\documentclass{aa}  

\usepackage{graphicx}
\usepackage{txfonts}
\begin{document}

   \title{Wavelength Calibration and Resolving Power\\ of the JWST MIRI Medium Resolution Spectrometer}


   \author{A. Labiano
          \inst{1,2},
          I. Argyriou\inst{3}
          \and
          J. \'Alvarez-M\'arquez\inst{4}
          \and
          A. Glasse\inst{5}
          \and
          A. Glauser\inst{6}
          \and
          P. Patapis\inst{6} 
          \and
          D. Law\inst{7} 
          \and
          B. R. Brandl$^{8,9}$ 
         \and
         K. Justtanont$^{10}$
         \and
         F. Lahuis$^{11}$
          \and
          J.R. Mart\'inez-Galarza$^{12}$
          \and
          M. Mueller$^{11}$
          \and
          A. Noriega-Crespo$^{7}$
         \and
         P. Royer$^{3}$
          \and
          B. Shaughnessy$^{13}$
         \and
         B. Vandenbussche$^{3}$
          }

   \institute{Centro de Astrobiolog\'ia (CSIC-INTA), ESAC - Camino Bajo del Castillo s/n, 28692 Villanueva de la Ca\~nada, Spain
             \and
             Telespazio UK for the European Space Agency (ESA), ESAC, Spain.
             \and
             Instituut voor Sterrenkunde, KU Leuven, Celestijnenlaan 200D, bus-2410, 3000 Leuven, Belgium.
            \and
            Centro de Astrobiolog\'ia (CSIC-INTA), Carretera de Ajalvir, 28850 Torrej\'on de Ardoz, Madrid, Spain.
            \and
            UK Astronomy Technology Centre, Royal Observatory, Blackford Hill, Edinburgh EH9 3HJ, UK.
            \and
            Institute of Particle Physics and Astrophysics, ETH Zurich, Wolfgang-Pauli-Str 27, 8093 Zurich, Switzerland.
            \and
            Space Telescope Science Institute, 3700 San Martin Drive, Baltimore, MD 21218, USA.
            \and
            Leiden Observatory, Leiden University, P.O. Box 9513, 2300 RA Leiden, The Netherlands.
            \and
            Faculty of Aerospace Engineering, Delft University of Technology, Kluyverweg 1, 2629 HS Delft, The Netherlands.
            \and
            Chalmers University of Technology, Onsala Space Observatory, S-439 92 Onsala, Sweden.
            \and
            SRON Netherlands Institute for Space Research, PO Box 800, 9700 AV, Groningen, The Netherlands.
            \and
            Center for Astrophysics, Harvard \& Smithsonian, 60 Garden St., Cambridge, MA,  02138, USA.
            \and
            Rutherford Appleton Laboratory, Harwell Campus, Didcot, Oxfordshire, OX11 0QX, UK.
             }

   \date{Received ..., 2021; accepted ..., 2021}

\newcommand{\mum}{$\mu$m}
\newcommand{\kms}{km s$^{-1}$}
\newcommand{\slice}{$_\mathrm{slice}$}
\newcommand{\mone}{$^{-1}$}
\newcommand{\mtwo}{$^{-2}$}
\newcommand{\isoa}{iso-$\alpha$}
\newcommand{\isol}{iso-$\lambda$}

  \abstract
   {The Mid-Infrared Instrument (MIRI) on-board the James Webb Space Telescope will provide imaging, coronagraphy, low-resolution spectroscopy and medium-resolution spectroscopy at unprecedented sensitivity levels in the mid-infrared wavelength range.The Medium-Resolution Spectrometer (MRS) of MIRI is an integral field spectrograph that provides diffraction-limited spectroscopy between 4.9 and 28.3 \mum, within a FOV varying from $\sim$13 to $\sim$56 arcsec square.
      {The design for MIRI MRS conforms with the goals of the JWST mission to observe high redshift galaxies and to study cosmology as well as observations of galactic objects, stellar and planetary systems. 
   }}
   {From ground testing, we calculate the physical parameters essential to general observers and calibrating the wavelength solution and resolving power of the MRS is critical for maximizing the scientific performance of the instrument.}
   {We have used ground-based observations of discrete spectral features in combination with Fabry-Perot etalon spectra to characterize the wavelength solution and spectral resolving power of the MRS.
   We present the methodology used to derive the MRS spectral characterization, which includes the precise wavelength coverage of each MRS sub-band, computation of the resolving power as a function of wavelength, and measuring { slice-dependent} spectral distortions.}
   {The ground calibration of the MRS shows that it will cover the wavelength ranges { from 4.9 to 28.3 \mum,} divided in 12 overlapping spectral sub-bands. The resolving power, is R$\gtrsim$3500 in channel 1, R$\gtrsim$3000 in channel 2, R$\gtrsim$2500 in channel 3, and R$\gtrsim$1500 in channel 4. 
   { The MRS spectral resolution optimises the sensitivity for detection of spectral features with a velocity width of $\sim$ 100 \kms\ which is characteristic of most astronomical phenomena JWST aims to study in the mid-IR}. 
   Based on the ground test data, the wavelength calibration accuracy is estimated to be below one tenth of a pixel (0.1 nm at 5 \mum ~and 0.4 at 28 \mum), with small systematic shifts due to the target position within a slice for unresolved sources, that have a maximum amplitude of about 0.25 spectral resolution elements.  The absolute wavelength calibration is presently uncertain at the level of 0.35 nm at 5 \mum\ and 46 nm at 28 \mum, and will be refined using in-flight commissioning observations.
   }
   {Based on ground test data, the MRS complies with the spectral requirements for both the R and wavelength accuracy for which it was designed.
   We also present the commissioning strategies and targets that will be followed to update the spectral characterisation of the MRS. }

    \titlerunning{JWST/MIRI MRS Spectral Characterization}

    \authorrunning{A. Labiano et al.}

   \maketitle
%

\section{Introduction}



The James Webb Space Telescope (JWST) is a space infrared observatory developed by the National Aeronautics and Space Administration (NASA), the European Space Agency (ESA) and the Canadian Space Agency (CSA). JWST comprises the Optical Telescope Element (OTE), an Integrated Science Module (ISIM), a Sun shield, and a Spacecraft Bus. The OTE collects and feeds the light to the scientific instruments in the ISIM. The main component of the OTE is the JWST segmented 6.5-meter primary mirror, made of 18 gold-plated beryllium, deployable, hexagonal mirrors. The ISIM is the structure that harbors the 4 scientific instruments of JWST: the Near Infrared Camera \citep[NIRCam,][] {nircam} the Near Infrared Spectrograph \citep[NIRSpec,][]{nirspec}, the Near Infrared Imager and Slitless Spectrograph \citep[NIRISS,][]{niriss} and the Mid-Infrared Instrument \citep[MIRI,][]{miri_pasp_1}. The JWST 5-layered sun shield is designed to block the light from the Sun, Earth and Moon, and allow the OTE and ISIM to passively cool to $\sim$40 K. MIRI then has a dedicated cryo-cooler to allow it to reach its operating temperature of about 7 K. The spacecraft bus carries all the ambient temperature equipment necessary to operate the JWST, and communicate with the ground stations. The expected launch date JWST is late 2021, and it will take about one month to reach its halo orbit around the second Earth-Sun Lagrangian point L2. It has a nominal mission time of 5 years, with a goal of 10 years. 

\begin{figure*}[h]
\centering
\includegraphics[width=2\columnwidth]{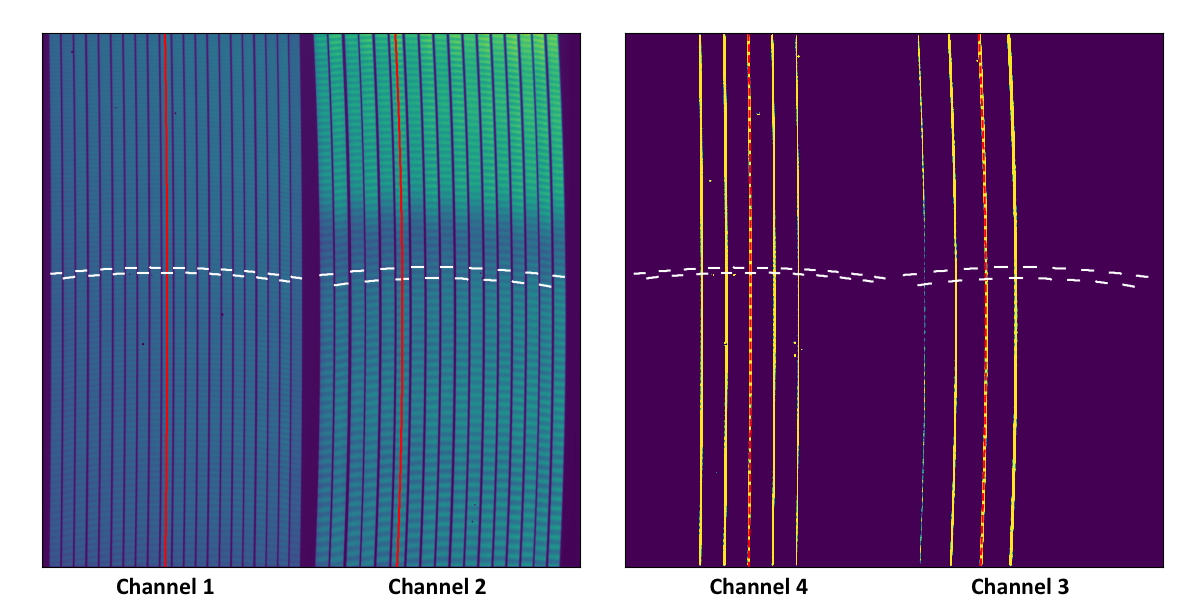}
\caption{Left: MRS SW detector (channels 1 and 2) illumination caused by an extended source filling the MRS FOV. 
Right: MRS SW detector illumination caused by a compact source. The illumination pattern along the slices follow the \isoa ~lines (see text for details). The \isoa ~and \isol ~lines are highlighted with red (\isoa) and white (\isol) lines overlaid on the detector image.}
\label{mrs_det}
\end{figure*}


With a spectral coverage from 4.9 to 28.3 \mum, MIRI is the only mid-infrared instrument on board JWST. It has four scientific operation modes: photometric imaging, coronagraphy, low resolution (R$\sim$100) long-slit spectroscopy, and medium resolution (R $\sim$1500 to 3500) integral field spectroscopy.  The imaging is performed by the MIRI Imager (MIRIM) in nine bands, within a field of view (FOV) of up to 2.3 square arcminutes \citep{Bouchet15}, depending on the subarray configuration selected. Coronagraphy is performed by three four-quadrant phase masks (4QPMs) at 10.65, 11.4, and 15.5 \mum, with a FOV of 24$\arcsec\times$24$\arcsec$, plus a Lyot coronagraph at 23 \mum\ with a FOV of 30$\arcsec\times$30$\arcsec$  \citep{Boccaletti15}. The Low Resolution Spectrometer is a long-slit (0.5$\arcsec\times$4.7$\arcsec$) spectrometer with both slit and slitless capabilities from 5 to 14 \mum~\citep{Kendrew15}. The Medium Resolution Spectrometer \citep[MRS,][]{Wells15} is an integral field spectrograph (IFS) that provides diffraction-limited spectroscopy between 4.9 and 28.3 \mum, within a FOV varying from $\sim$12 to $\sim$56 arcsec square.


MIRI will provide about 50 times the sensitivity and 7 times the angular resolution of Spitzer. It is expected to make important contributions to all four of the JWST science pillars \citep[first light of the Universe, assembly of galaxies, star and planet formation, and evolution of planetary systems plus conditions for life, e.g. ][]{miri_pasp_1} and astronomy in general. In fact, 41$\%$ of the submitted Cycle 1 General Observer proposals requested to use MIRI, and $31\%$ had MIRI as the Prime Instrument Mode. Half of them ($14.5\%$) requested the MIRI MRS as Prime\footnote{https://www.stsci.edu/jwst/science-execution/approved-programs/cycle-1-go}.
{ 
The MRS was designed to provide medium resolution (R $= \lambda/\Delta\lambda > 1000$) 3D spectroscopy in the whole MIRI range. It will be able to resolve (or separate) lines that have a velocity widths (separations) of about 100 \kms. 
The scientific goals behind this requirement include \citep[e.g.,][]{Gardner06}: 
to study the evolution of the Hydrogen (Lyman, Balmer, Paschen) lines ratios and derive the ionizing continuum of galaxies and ionization state of the Universe as a function of redshift (this measurement requires R = 1000 spectroscopy of of the Hydrogen lines, with intensities down to $2\times 10^{-19}$ erg cm\mtwo s\mone); 
 to measure the relative motions between (and within) galaxies, and  scaling laws of their basic properties (luminosity, size, kinematics, metallicity, etc) from redshifts 1 to 7, revealing the physical processes responsible for those; 
 to search for the redshifts and power sources of the high redshift Ultra Luminous Infrared Galaxies using narrow emission lines such as [NeVI] 7.66 \mum\ or rest-frame optical lines at high redshift, and comparing them with the PAH emission (e.g., a R $>$ 1000 will allow separating the [NII] and H$_\alpha$ lines, plus detection and in some cases resolution of the [NeVI] 7.66\mum\ and PAH lines up to redshifts z$\sim$6, based on the extrapolation of the Arp220 spectra and Circinus galaxy, at 10$\sigma$, in less than 10s);
 to map and detect the molecular gas in galaxies through e.g., the H2 line at 28.22 \mum\ with adequate contrast of the emission line relative to the continuum (high spectral resolution is essential to obtain an  adequate contrast of the line over the continuum), where we expect the largest contributions from telescope and zodiacal backgrounds;
 to study, map, and trace the evolution of the physical properties, composition, and structure of young stars, brown dwarfs, faint companions, circumstellar disks, and protoplanets immediately adjacent to their much brighter neighbors (to detect a 10 $\mu$ Jy source, such as a typical Class 0 protostar in Taurus-Auriga, we need a R = 2000 at 15 \mum, plus a continuum sensitivity of$\sim7\times 10^{22}$ W m\mtwo\ per resolution element);
 to study and characterise the atmospheres and properties of exoplanets, combined with direct imaging of these (e.g., a 7 M$_\mathrm{JUP}$ object is bright enough that spectra at R $\sim$ 3000 can be collected, enabling atmospheric structure and composition to be inferred); 
 to resolve the spectral signatures of key ices and silicates, both in crystalline and amorphous form, in the disks and observe planets perturbing the disk structures (R = 3000 allows the detection of broad features such as absorption bands against the spectra of background sources, or emission tracers from warm material in the inner regions of disks), and study the mineralogy of the dust in nearby debris disks and of their precursors around young stars; 
 to determine the basic physical parameters of a brown dwarf atmosphere such as gravity, composition, the temperature-pressure profiles (gradient, inversions), mass, and the effect of clouds (a R $\sim$ 1000 is required to resolve most of the spectral features tracing the atmosphere properties);   
 to study the gas phase processes in the inner comae of active comets, revealing their composition (comets will have extremely rich spectra, with up to thousands of overlapping lines, thus JWST spectroscopy of comets requires R up to 3000 for some isotopic ratios); 
 to extend spectroscopic studies from the best ground-based studies of Pluto and Triton to more distant Kuiper Belt objects (a R = 3000 will allow to seek isotopic ratios in the water ice and other components, as well as monitor surface temperature through the nitrogen overtone band), provide compositional and isotopic data, and even separate the size and albedo of such objects, thereby constraining their physical size.}


The MRS consists of two optical modules: the Spectrometer Pre-Optics (SPO) and the Spectrometer Main Optics (SMO), plus the MRS internal calibration source. The SPO carries two dichroic and grating wheels assemblies (DGA) and a set of mirrors that split the light entering the MRS into 4 spectral channels, each one with a different optical path. The DGA dichroics also divide the light into three spectral sub-bands per channel. The SPO then feeds the light of the four channels into four dedicated integral field units (IFUs), each of which produces a 2D spectrum of the FOV. These spectra are recorded by two detectors: one for the short wavelength channels (SW detector), one for the long wavelength channels (LW detector).

The four spectral channels of the MRS are named 1, 2, 3, and 4 in sequential order from the shortest wavelength to the longest, 
each consistent of sub-bands SHORT (or A), MEDIUM (or B), and LONG (or C), from short to long wavelengths in each channel { (see Table \ref{spectab})}. The fields of view of the four channels are roughly concentric, with the smaller FOV corresponding to channel 1 and the largest to channel 4. Hence, to obtain a spectrum of a target for the whole MRS wavelength range, the observer needs to stitch together 12 sub-band spectra, obtained in three different exposures. The first exposure would cover the 1A-2A-3A-4A bands, the second one bands 1B-2B-3B-4B, and the third one bands 1C-2C-3C-4C. 
By default, the JWST pipeline will produce a spectral cube (two axes with spatial information plus a third axis with the spectral information) per MRS band, although it is also capable of producing a spectral "super-cube" in which all 12 bands are combined \citep[see, e.g., early descriptions by][]{MRSpipeline,gordon15}.



Each IFU of the MRS divides its FOV in a different number of slices whose widths (0.176$\arcsec$, 0.277$\arcsec$, 0.387$\arcsec$, 0.645$\arcsec$ from channel 1 to channel 4) are designed to be the same fraction (roughly the FWHM) of the diffraction-limited PSF for all channels. Thus, the channel 1 IFU contains 21 slices (FOV = $3.2\arcsec\times3.7\arcsec$), channel 2 (FOV = $4.0\arcsec\times4.8\arcsec$) has 17 slices, 16 slices for channel 3 (FOV = $5.5\arcsec\times6.2\arcsec$), and 12 slices for channel 4 (FOV = $6.9\arcsec\times7.9\arcsec$). The IFUs of channels 1 and 2 disperse their light on the SW detector, while channels 3 and 4 are dispersed on the LW detector. All slices of a single IFU cover roughly one half of a detector. Figure \ref{mrs_det} shows the MRS detectors illuminated by an extended (left panel) and point source (right panel), which fills the full FOV of the four channels.  A full description of the MRS design and operations can be found in \citet{Wells15,pasp_det}. \citet{Yannis20} and \cite{Glauser10} include a detailed description of the MRS FOV coordinate system, and the detector -- sky projection and reconstruction. 

Within each channel FOV, we use two coordinates to define the spatial directions: the $\alpha$-axis is defined parallel to the IFU image slicer along-slice direction, the $\beta$-axis is defined perpendicular to the $\alpha$-axis and describes the slicer across-slice direction. Due to the optical design of the MRS, the region illuminated by each slice on the detector follows a curved line; the spatial ({\it$\alpha$}) and spectral ({\it$\lambda$}) directions of the dispersed spectrum are not parallel to the x and y coordinate directions of the detector pixels. Within a single slice, we define as {\it isoalpha} (or \isoa) the lines on the detector where the $\alpha$ coordinate remains constant. Thus, an \isoa line traces constant spatial position at various wavelengths. In the same way, {\it isolambda} (or \isol) lines are defined by the coordinates of constant wavelength on the detector. The data from the MIRI test campaigns show that the \isoa lines are well described by third order polynomials while the \isol lines are described by second order polynomials. Thus, both the {\isoa} and {\isol} are functions of (x,y) detector coordinates, with a unique and discontinuous function for each slice (Patapis et al. in prep.).
An example of {\isoa} and {\isol} lines on the detector are shown in Fig. \ref{mrs_det}.


In this contribution, we discuss the methods used to calibrate the wavelength solution
and spectral resolving power of the MRS using ground test data. Unlike at optical
wavelengths,
in the mid-infrared there are no arc lamps available that can provide
a well sampled grid of bright, unresolved emission lines throughout the wavelength range
of the MRS.  We therefore use a hybrid approach that combines Zemax optical models, Fabry-Perot etalons, and a variety of absolute wavelength reference points
in order to derive a comprehensive wavelength solution.


In \S \ref{obs.sec} we describe the ground-test data that forms the backbone of the MRS
wavelength solution, in particular the Fabry-Perot etalons that produce an
interference pattern with a well-defined peak-to-peak separation.
We then describe how we define a small number of absolute wavelength
reference points in each of the twelve MRS bands using features in the spectral
response function of a variety of filters.  By combining these reference points
with the Fabry-Perot data, we demonstrate how we are able to expand the wavelength
solution to cover the entire MRS spectral range in \S \ref{relative.sec}, with
additional adjustments to the relative solutions in each detector pixel using
super-sampled observations of the line profiles.  We expand upon this analysis
to estimate the spectral resolving power in \S \ref{resolution.sec}.
We note a few caveats to this ground-based solution in \S \ref{caveats.sec}, and conclude
with a discussion in \S \ref{flight.sec} of our strategy for future refinement of our
solutions using in-flight observations.
We summarize our conclusions in \S \ref{summary.sec}.





\section{Ground-Test Observations}
\label{obs.sec}

Before its delivery to NASA in 2012, MIRI underwent two ground test campaigns: one for the verification model (VM), and one for the flight model (FM). The VM was a fully operational version of MIRI, with reduced scientific functionalities,  built as a pathfinder for the test and calibration campaigns, to de-risk the opto-mechanical concepts and assembly integration and verification program. The main goal of the VM campaign was to verify the instrument optical performance in realistic conditions to detect possible problems that could affect the FM. It also served as a test bed for the FM campaign planning and test designs \citep{miri_pasp_2, FMpaper, Glasse10, VMpaper}. The FM campaign was much more extensive than the VM campaign, and aimed to 
more fully characterize the performance of MIRI under flight conditions. All the data relevant for the spectral characterization of the MRS were obtained during the FM campaign, and were processed through the standard MIRI ramps-to-slopes routines, background subtraction and flat-field correction of the resulting slope images. No fringe flats were applied since not all fringe flats were available by the time of the analysis. Our tests show that at the signal to noise ratio (SNR) obtained, not including the fringe flats does not affect the measured centroids of the etalon lines by more than a few percent of the resolution element \citep{Yannis20}. Throughout this paper we will refer to these measurements as {\it FM results or FM data}.

\begin{figure}[t]
\centering
\includegraphics[width=\columnwidth]{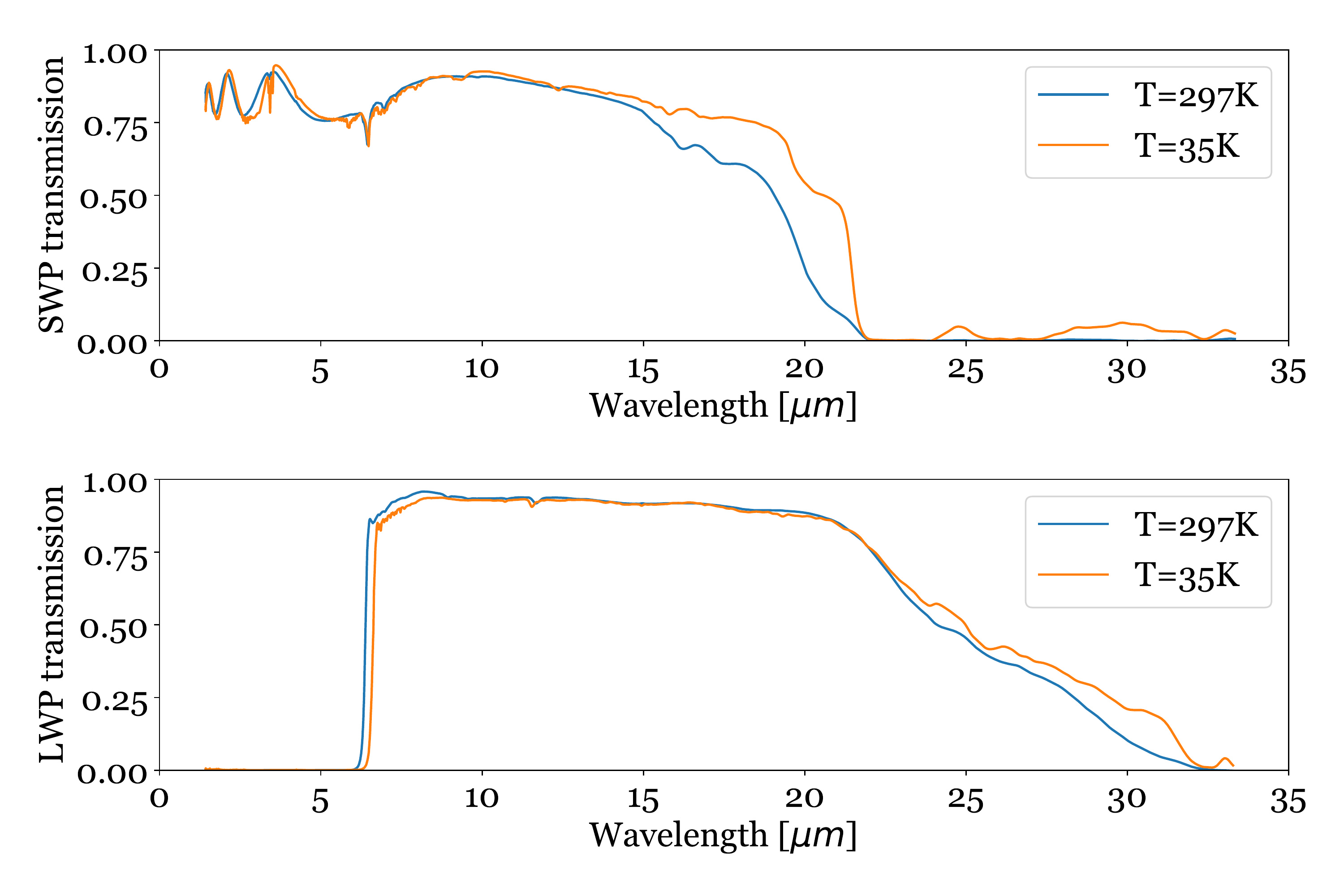}
\caption{Transmission curves for the SWP filter (top) and the LWP filter, as measured in the laboratory by INTA, before mounting them on the MTS filter wheel.  Sharp spectral features in the transmission curves are used
for defining absolute wavelength reference points for
the MRS.}
\label{transmissions}
\end{figure}

To recreate the MIRI input signal from JWST, both campaigns used the MIRI Telescope Simulator \citep[MTS,][]{mts_2004,Herrada2007, mts_paper}. The MTS was an infrared cryogenic telescope simulator developed by the Instituto Nacional de T\'ecnica Aeroespacial (INTA) that delivered a diffraction-limited illumination beam reproducing the main particularities introduced by the JWST optics, including both point source and extended source illumination capabilities. The MTS contained a filter wheel which included four Fabry-Perot etalons and two wave-pass filters.
The etalons were designed to provide unresolved spectral features across the entire MIRI wavelength range\footnote{Each of the four etalons was optimized for a single MRS channel. For convenience, we will name them: ET1 for Channel 1, ET2 for Channel 2, ET3 for channel 3, and ET4 for Channel 4.},
while the long (LWP) and short (SWP) wave-pass filters 
have sharp cut-offs at 6.6 and 21.6 \mum\ respectively (see Fig. \ref{transmissions}) that provide
well-defined wavelength reference points at a temperature
of 35 K.

\begin{figure}[t]
\includegraphics[width=0.95\columnwidth]{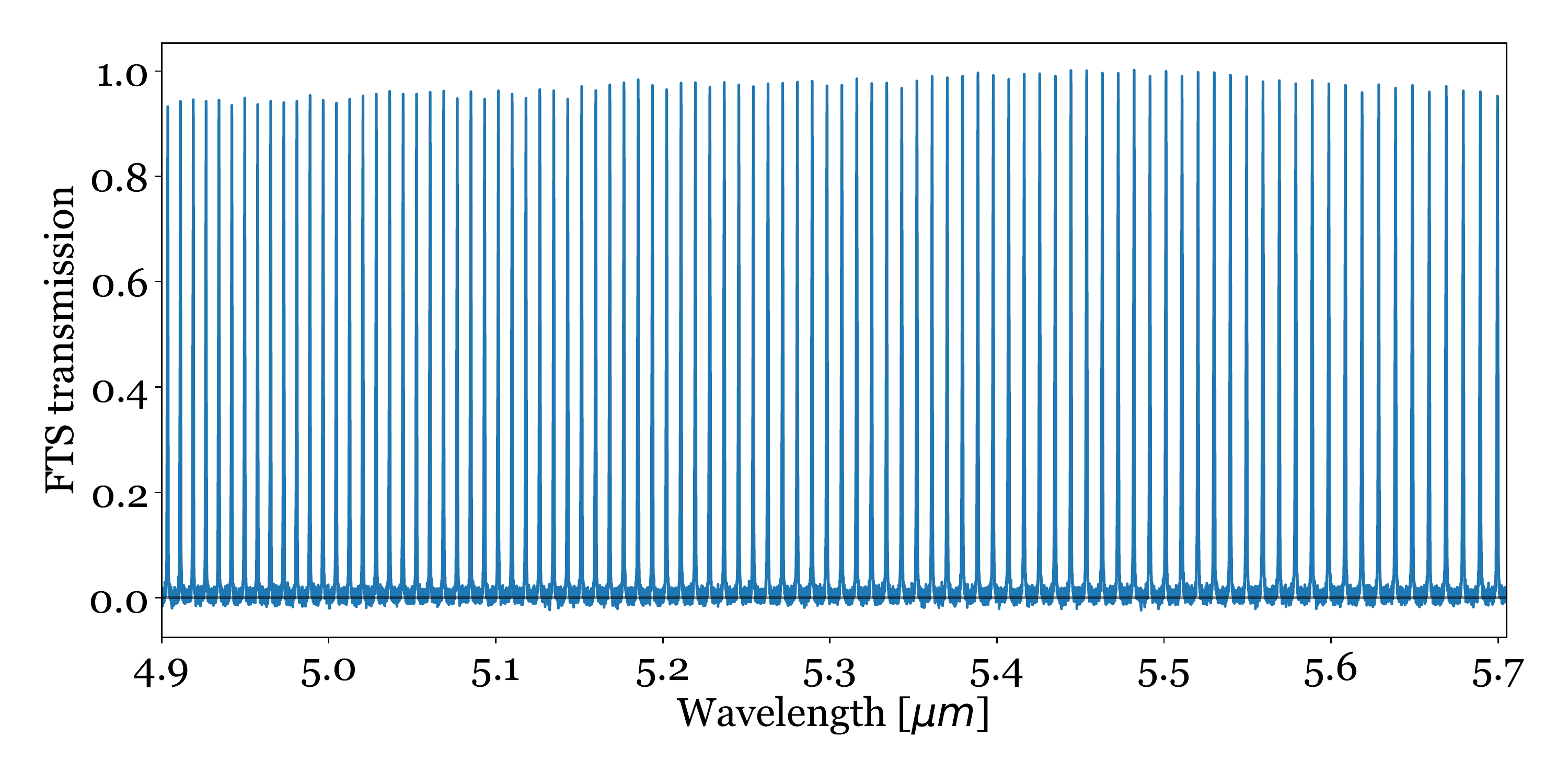} 
\includegraphics[width=\columnwidth ]{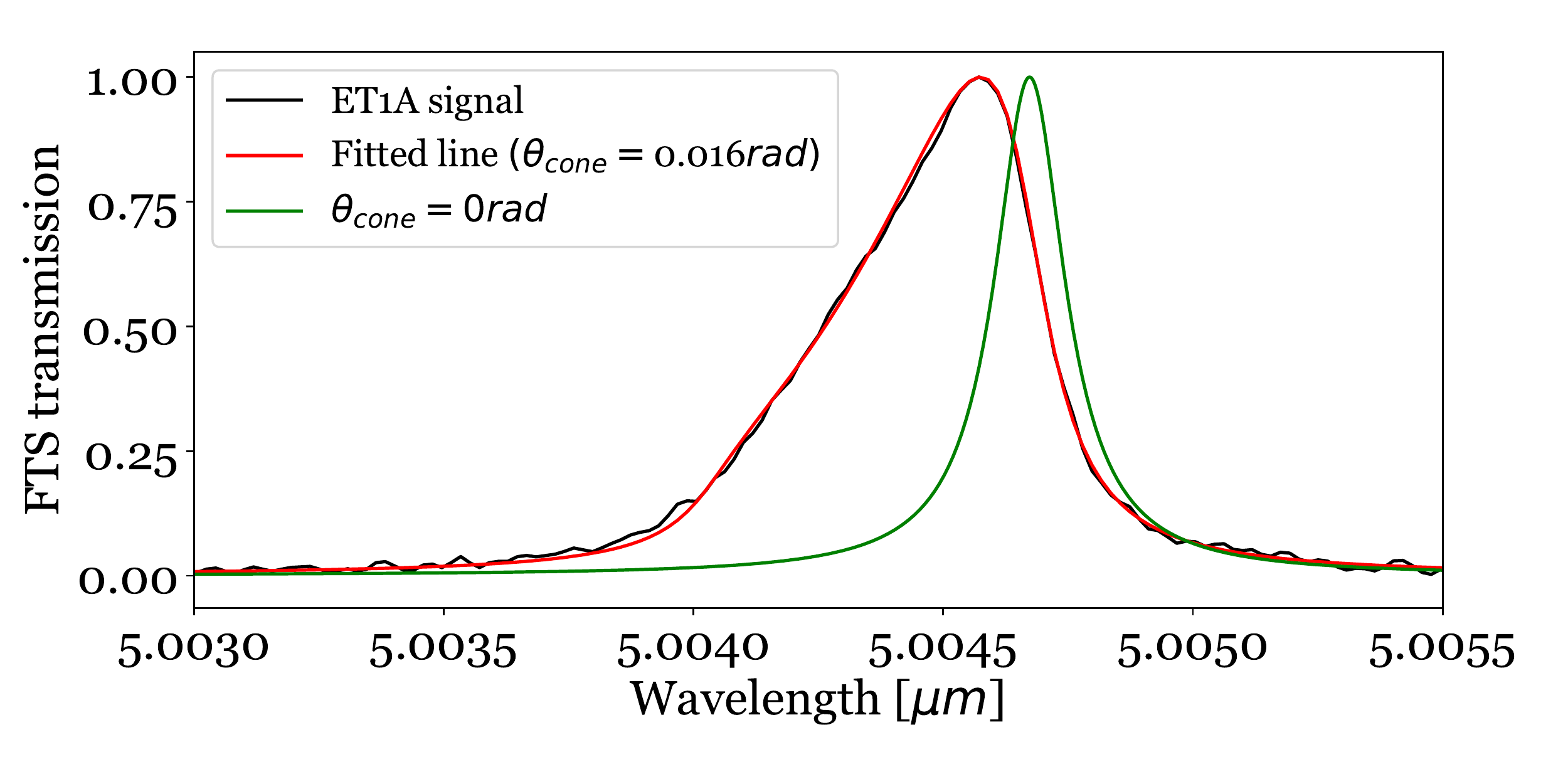} 
\includegraphics[width=0.95\columnwidth]{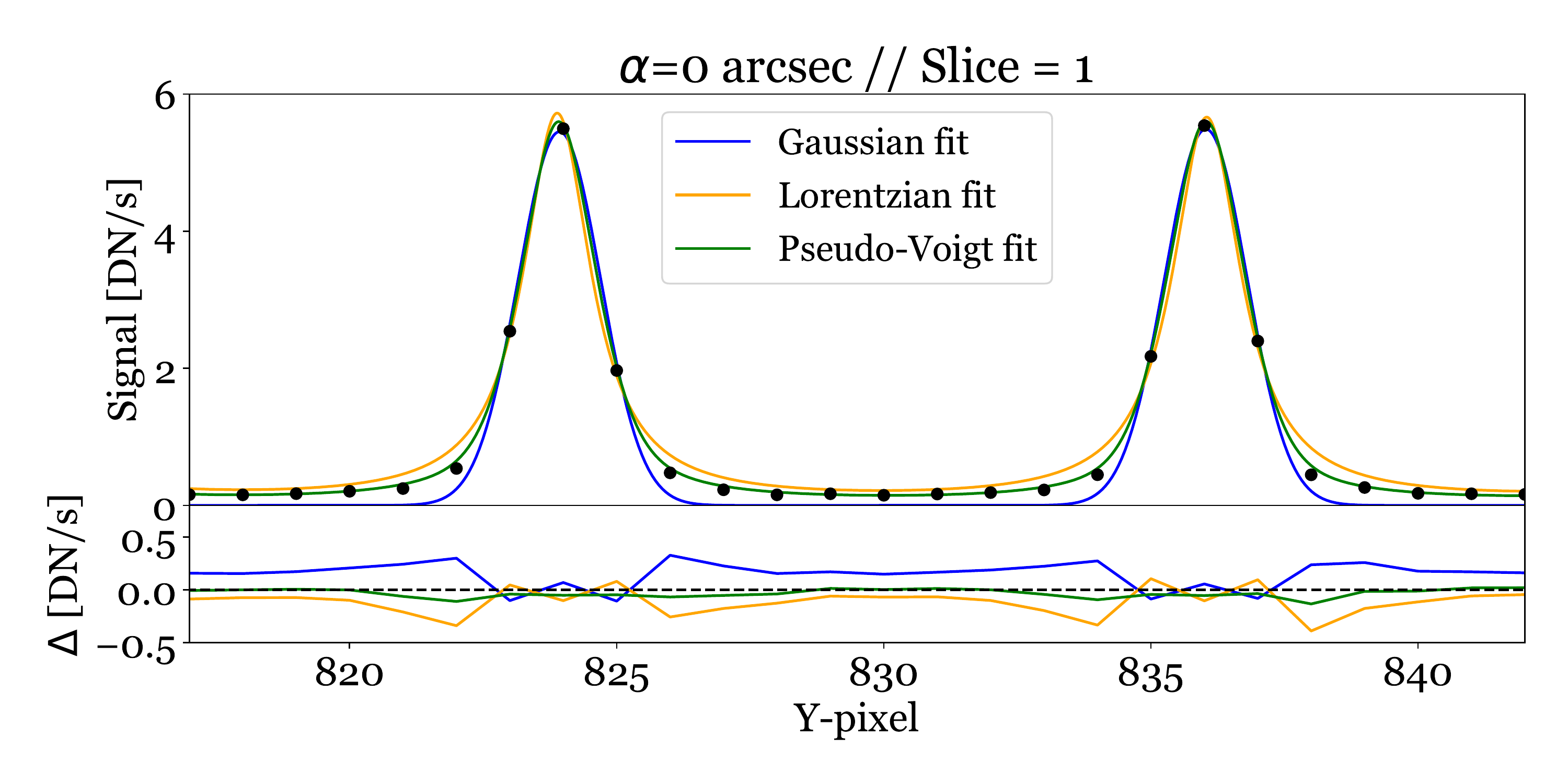}
\caption{
Top: transmission of ET1 measured with the FTS for the wavelength range of MRS band 1A (FTS data), before mounting it on the MTS for the MIRI test campaigns. Multiple well-separated (unblended) resonance lines can be seen.
Middle: fit of a single etalon line in the ET1 FTS data (black profile). The line shape in the data is impacted by the non-zero divergence angle of the beam (deviation from a perfectly collimated beam), and shows up as here as the asymmetry of the line. We account for this in the fit (see Sect. \S \ref{obs.sec}, as shown by the red profile. The green profile shows the same etalon line if the beam was perfectly collimated (Eqs.~\ref{eq:norm_factor} and \ref{eq:transm_function_solid_angle}). 
Bottom: fit of different line profiles from the (MTS mounted) ET1 etalon, as observed with the MRS (black points), and residuals of the fits computed by subtracting the fitted model from the data. The MTS+MIRI setup is not affected by collimation issues. Its spectral resolution, as expected, is clearly lower than in the FTS measurements.
}
\label{ET1A}
\end{figure}

\begin{table}[t]
\caption{\label{spectab} Spectral characteristics of the MRS from Zemax modeling.}
\centering
\begin{tabular}{cccccccccc}
\hline\hline
MRS    &MRS &Nominal coverage$^*$ & Pixel scale $^**$\\
Channel&Band& (\mum) & \mum/pixel \\ 
\hline
   & SHORT (A)    &   4.87 --  5.81 & 0.0009 \\
1 & MEDIUM (B)  &   5.62 --  6.73 & 0.0011  \\
    & LONG (C)     &   6.49 --  7.76 & 0.0012\\
\hline
  & SHORT (A)    &   7.45 --  8.90 & 0.0014 \\
2 & MEDIUM (B)  &   8.61 -- 10.28 &  0.0016 \\
    & LONG (C)     &   9.91 -- 11.87 & 0.0019 \\
\hline
  & SHORT (A)    &  11.47 -- 13.67 & 0.0021 \\
3 & MEDIUM (B)  &  13.25 -- 15.80 &  0.0025\\
    & LONG (C)     &  15.30 -- 18.24 & 0.0029 \\
\hline
  & SHORT (A)    &  17.54 -- 21.10 & 0.0035 \\
4 & MEDIUM (B)  &  20.44 -- 24.72 & 0.0042  \\
    & LONG (C)     &  23.84 -- 28.82 & 0.0049 \\
\hline
\end{tabular}
\tablefoot{\tablefoottext{*}{These ranges correspond to the Zemax model results, not the final, wavelength calibrated ranges \citep{Wells15}.}\tablefoottext{**}{Nominal wavelength coverage / 1024 pixels.} The full details of the MRS characteristics can be found in \cite{Wells15} and JDox pages.}
\end{table}
\begin{table*}[t]
\caption{\label{RPs} Location of relevant features used for MRS spectral characterisation.}
\centering
\begin{tabular}{cccccccccc}
\hline\hline
 & & \multicolumn{4}{c}{Relative wavelength calibrators} & \multicolumn{3}{c}{Absolute wavelength calibrators (RPT)} \\ 
MRS    &MRS & ET1 & ET2 & ET3 & ET4 & LWP & SWP & X-Dichroic \\
Channel&Band& lines & lines & lines & lines & features & features & features \\
\hline
1 & SHORT (A)     & X & X \\
1 & MEDIUM (B)   & X & X &  &  &  & X \\
1 &  LONG (C)      & X & X &  &  & X \\
\hline
2 & SHORT (A)     & X & X &  &  &   & & X \\
2 & MEDIUM (B)  & X & X & X &  &   & & X \\
2 &  LONG (C)      & X & X &  X & & X \\
\hline
3 & SHORT (A)     & X & X & X \\
3 & MEDIUM (B)   & X & X & X & X  \\
3 &   LONG (C)   & X & X & X & X \\
\hline
4 & SHORT (A)   & X &  & X & X & & & X \\
4 & MEDIUM (B)  &   &   &   & X &   & X & X  \\
4 &  LONG (C)    &   &   &   & X &   &   & X  \\
\hline
\end{tabular}
\end{table*}
%


The pure transmission (with no MTS or MIRI involved) of the etalons at cryogenic temperatures ($\sim$80 K) was measured at Rutherford Appleton Laboratory (RAL) using a Fourier Transform Spectrograph (FTS). Each etalon was taken out of the MTS and mounted on a block cooled with continuously flowing liquid nitrogen, and illuminated with quasi-collimated beam (divergence angle $<1\deg$) at normal incidence. Throughout the paper we will refer to these measurements as {\it FTS data}. The uncertainties associated to these measurements are not available.
Figure \ref{ET1A} shows the transmission curve of ET1, as an example of the etalon FTS data, in the MRS band 1A wavelength range. Constructive interference within the etalon layers results in clear transmission peaks, whilst destructive interference results in no line appearing in the spectrum.

The middle panel of Fig. \ref{ET1A} shows a zoom into one etalon  line ET1. There is a prominent skewness of the line, which substantially complicates the determination of both the wavelength of the line center and its FWHM. This skewness is in fact present in all the lines, and it is due to the finite diameter of the FTS pupil giving a cone angle of a few degrees at each point on the etalon. What we observe is our reference line integrated over a cone angle (range of incidence angles). Given that the requirement on the spectral calibration accuracy is 10\% of a resolution element\footnote{The 1/10th resolution element calibration accuracy was driven by the acceptance that mapping velocity structure below 30 km/s at more than a single grating setting was not a goal.  Mass and space constraints required MIRI/MRS to break the spectrum into bands which were selected using moving mechanisms.  1/10th spectral resolution element was then seen as a credible design goal, with the mechanisms designed specifically to provide optimum stability and reproducibility in the wavelength dimension.}, we need to account for the cone angle variation. 

Mathematically, the etalon line model is then described by the following set of equations. Equation \ref{eq:finesse_FTS} defines the "coefficient of finesse", a quantity linked to the reflectivity of an optical material. The higher the reflectivity, the higher the coefficient of finesse, and the larger the contrast of an etalon line. Equation \ref{eq:delta_FTS} defines the phase difference between successive emission peaks. We use Eq.~\ref{eq:finesse_FTS} and Eq.~\ref{eq:delta_FTS} in the Fabry-P\'erot transmittance function given by Eq.~\ref{eq:transm_function_FTS} \citep{lipson1969}. This function describes the transmission spectrum of an etalon based on its reflectivity, the refractive index of the material used, and the optical thickness of the etalon as a function of wavelength and incidence angle.

\begin{equation}\label{eq:finesse_FTS}
F_R(\lambda ') = \frac{\pi \sqrt{R_{\lambda} (\lambda ')}}{1 - R_{\lambda} (\lambda ')}
\end{equation}

\begin{equation}\label{eq:delta_FTS}
\delta(\lambda ',\theta) = \frac{4\pi \cdot n_{R} \cdot D_{gap} \cdot \cos(\theta)}{\lambda '}
\end{equation}

\begin{equation}\label{eq:transm_function_FTS}
    T_\theta (\lambda ', \theta) = \frac{1}{1 + \left(\frac{2F_{R}(\lambda ')}{\pi} \right)^2 \cdot \sin^2\left(\frac{\delta(\lambda ',\theta)}{2} \right)}
\end{equation}

The above set of equations describe an etalon line produced by a ray, yielding a symmetric line. If the etalon is illuminated by rays with a finite cone angle, Eq.~\ref{eq:transm_function_FTS} has to be integrated over the profile of the incident light. To do so we multiply Eq.~\ref{eq:transm_function_FTS} by Eq.~\ref{eq:gaussian_solid_angle} (where we assume an intensity which has a Gaussian dependence on angle of incidence on the etalon), and then we integrate over all incidence angles. As a last step we normalize the line profile, as shown by Eq.~\ref{eq:norm_factor} and \ref{eq:transm_function_solid_angle}.

\begin{equation}\label{eq:gaussian_solid_angle}
f(\theta) =  2\cdot \pi \cdot \theta \cdot \exp^{-\left(\frac{\theta}{\theta_{cone}}\right)^2}
\end{equation}

\begin{equation}\label{eq:norm_factor}
f_{norm}(\theta_{cone}) =  \int^{\theta_{cone}}_{\theta_{off}} f(\theta ') d\theta '
\end{equation}

\begin{equation}\label{eq:transm_function_solid_angle}
    T_\lambda (\lambda ', \theta_{cone}) = \frac{\int^{\theta_{cone}}_{\theta_{off}} f(\theta ') \cdot T_{\theta}(\lambda ',\theta ')d\theta '}{f_{norm}(\theta_{cone})}
\end{equation}

We find a best fit to the observed etalon profile using $\theta_{cone}$ = 0.016 radians, as illustrated in Fig. \ref{ET1A}, where the offset between the centroids of the best fit (red) and the intrinsic etalon profile (green) is found to be $10^{-5} \mu m$. 
This translates to an offset of about a tenth of a pixel in MRS band 1A. 
Hence, we fitted the collimated beam line profiles to obtain the line wavelengths and FWHM for the 4 etalons.

An (MTS-mounted) etalon line, as observed with the MIRI MRS, is a convolution between the intrinsic etalon line profile (take for example the green curve in Fig.~\ref{ET1A}) and a Gaussian profile with a width defined by the spectral resolution of the MRS. An etalon line profile is given by Eq.~\ref{eq:transm_function_FTS}. This line profile is mathematically equivalent to a Lorentzian line profile. Convolving a Lorentzian line profile with a Gaussian line profile yields a Voigt profile. Due to the limited number of data points in the MRS data, to fit the FM etalon lines we define a "pseudo-Voigt" profile by taking the fractional contributions of a Gaussian and a Lorentzian line shape and summing them. The process is portrayed by Eq.~\ref{eq:gaussian_profile}, \ref{eq:lorentzian_profile}, and \ref{eq:voigt_profile}.

\begin{equation}\label{eq:gaussian_profile}
    G(\mu,\sigma) = \frac{1}{\sigma} \cdot \sqrt{\frac{4~ln(2)}{\pi}} \cdot e^{-4~ln(2)\left(\frac{z - \mu}{\sigma}\right)^2} 
\end{equation}

\begin{equation}\label{eq:lorentzian_profile}
    L(\mu,\sigma)   = \frac{2}{\pi\sigma} \cdot \frac{1}{1 + 4 \left( \frac{z - \mu}{\sigma}\right)^2}
\end{equation}

\begin{equation}\label{eq:voigt_profile}
    V(A,\mu,\sigma) = A \left[f \cdot L(\mu,\sigma) + (1 - f) \cdot G(\mu,\sigma)\right]
\end{equation}

In the above expressions, $z$ can be wavelength or pixels, and the free-varying parameters are A, $\mu$, $\sigma$, and $f$. In the case where $f$ = 0 the above formalism reduces to a Gaussian line profile approximation. Alternatively, in the case where $f$ = 1 the above formalism reduces to a Lorentzian line profile approximation. 

\begin{figure*}
\centering
\includegraphics[width=180mm]{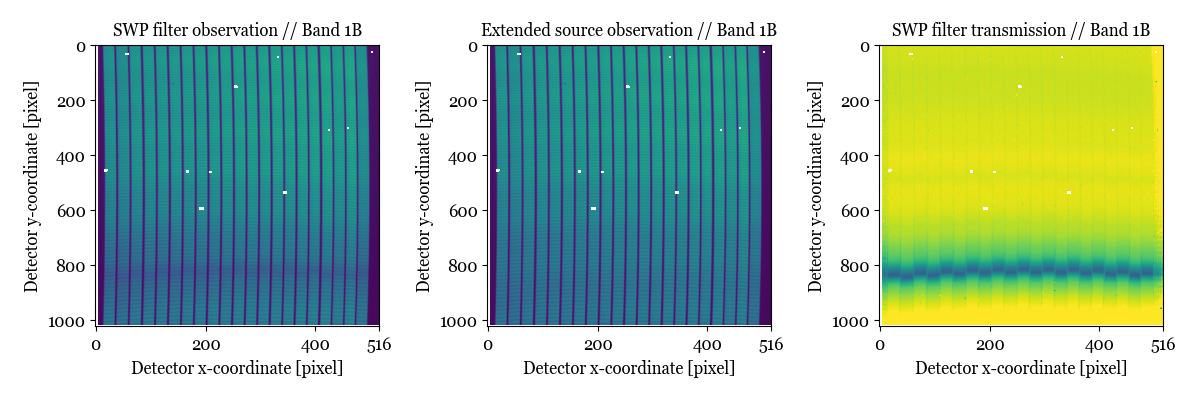}\\
\includegraphics[width=180mm]{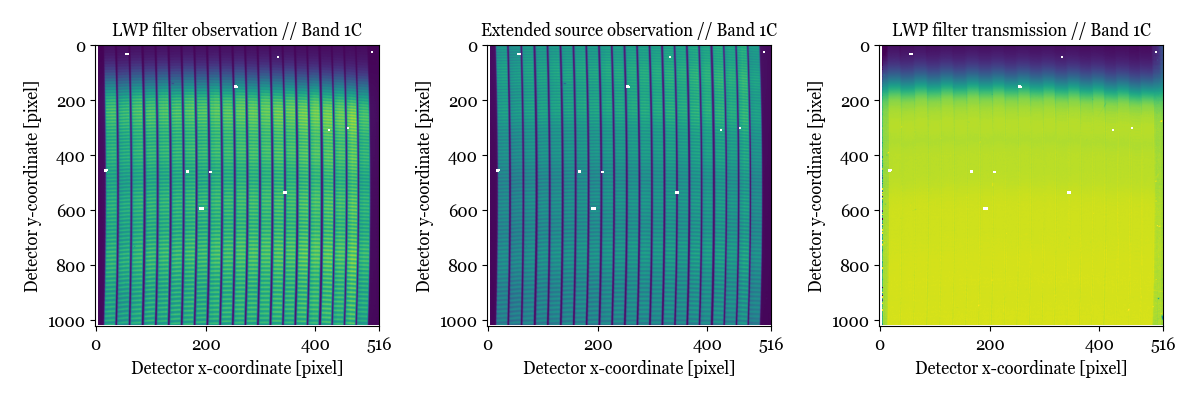}\\
\caption{From left to right: observation of the MTS black body emission through the SWP (top) and LWP (bottom) filters, and with no MTS filter selected (center). The right panels show the division of the first (left) image by the second (center) image, producing a detector plane image of the wave pass filter throughput curves. 
}
\label{fig_wp_det}
\end{figure*}

In Fig.~\ref{ET1A} we show the result of fitting an FM etalon spectrum, extracted from an \isoa line centered in a slice on the detector. The fitting is done with three different line profiles, a Gaussian, a Lorentzian, and a pseudo-Voigt profile. Each spectral line is fitted separately with each model, and then the fitted profiles are plotted together. The residuals of the fit of each model are shown in the bottom plot. We note that the data have been background-subtracted. Examining the residuals, we clearly see that the pseudo-Voigt model fits the line core and line wings best. The Gaussian model underestimates the line wing contribution to the lines, while the Lorentzian model overestimates it.


Having accurately determined the etalon line centroids in the FM data in pixel space, and the FTS data in wavelength space, the next step is to match the two sets of centroids.

\section{Absolute Calibration}
\label{absolute.sec}

The MTS etalons provide a plethora of unresolved, identifiable spectral features for the wavelength calibration of the MRS. However, the etalon lines are largely indistinguishable from one another so we cannot tell which is which when observing a line in the detector pixel space of the MRS without an external, absolute wavelength reference. Furthermore, a number of factors make the matching of the etalon lines of the FTS and FM data extremely difficult:

\begin{itemize}
    \item The spectrophotometric response curves of the instruments used (FTS and MTS+MIRI/MRS) are different.
    \item The optical chain before the etalons in the FTS bench and the MTS bench were not the same. As described in the previous section, we accounted for this by estimating the chief ray collimated line centroid in the FTS case, and assumed (based on the MTS design and the ground test data) that the MTS yielded a perfectly collimated beam.
    \item The FTS and FM data were taken at different temperatures (80 K for the FTS versus 35 K for the MTS), causing the refractive index of the etalon internal surfaces and the thermal coefficients of the material (which are temperature dependent) to be different during the FTS and FM measurements. Even though the effect is not expected to be large at these temperatures, it adds an additional, small uncertainty on the FTS-FM line centroids transformation.
\end{itemize}

To resolve these issues, an unambiguous reference point (RPT) is needed that matches
a known wavelength determined from the high-resolution FTS data with an (x,y) pixel
coordinate observed on the MRS detector.  This reference point can be used to identify
the nearest etalon line, and by extrapolation all of the other etalon lines on
the detector.
The FM campaign included several such tools that can be used to get the needed RPTs for the wavelength characterisation of most of the MRS spectral sub-bands.
As described in \S \ref{wavepass.sec} - \ref{overlap.sec}
(and Table \ref{RPs}), in order of preference these include spectral features in specially-constructed wave-pass filters
(used for Channels 1B, 1C, 2C, and 4B), features observed in the MRS order-selection
dichroic filters when in a crossed-configuration (used for Channels 2A, 2B, 4A, 4B, and 4C), and spectral overlaps between adjacent MRS bands (used for Channels 1A, 3A, 3B, and 3C). 
In this Section, we describe how we obtained the RPT for one pixel (x,y) coordinates on the central \isoa of each slice, for each channel. Section 4 shows how these RPTs were extrapolated to the rest of \isoa lines and slices in the channel.


\subsection{Spectral calibration reference points}
\label{wavepass.sec}

As described in Section \S \ref{obs.sec}, the MTS filter wheel included two wave-pass filters. The transmission of the wave-pass filters was measured in a laboratory at the University of Reading (henceforth UoR). The measurements were performed at room temperature and at 35~K and in the latter case an average uncertainty of $\pm$2\% was reported for the transmission values across the entire wavelength range. No uncertainties were reported on the wavelength measurements. 
One wave-pass filter was designed to give a RPT for the short wavelengths (SWP) and has a sharp cut-off at 21.6 \mum. The second wave-pass filter was built to produce a RPT for the long wavelengths (LWP), and has a sharp cut-off at 6.6 \mum. Both filters also show spectral features in other wavelengts that can be used as RPTs (see Table \ref{RPs}). Measurements of the wave-pass filters with the MRS were performed at 7~K using the MTS filter wheel during the FM campaign. To determine the filter transmission, mounted on the MTS under flight conditions, the MRS was first flood-illuminated using the MTS calibration source at 800~K. Thereafter the same measurement was performed by placing the wave-pass filter in the beam path. By dividing the measurements with and without the filter, we compute the filter transmission in the detector plane. The result is shown in Fig.~\ref{fig_wp_det} for MRS bands 1B and 1C as an example.

\begin{figure}
\centering
\includegraphics[width=90mm]{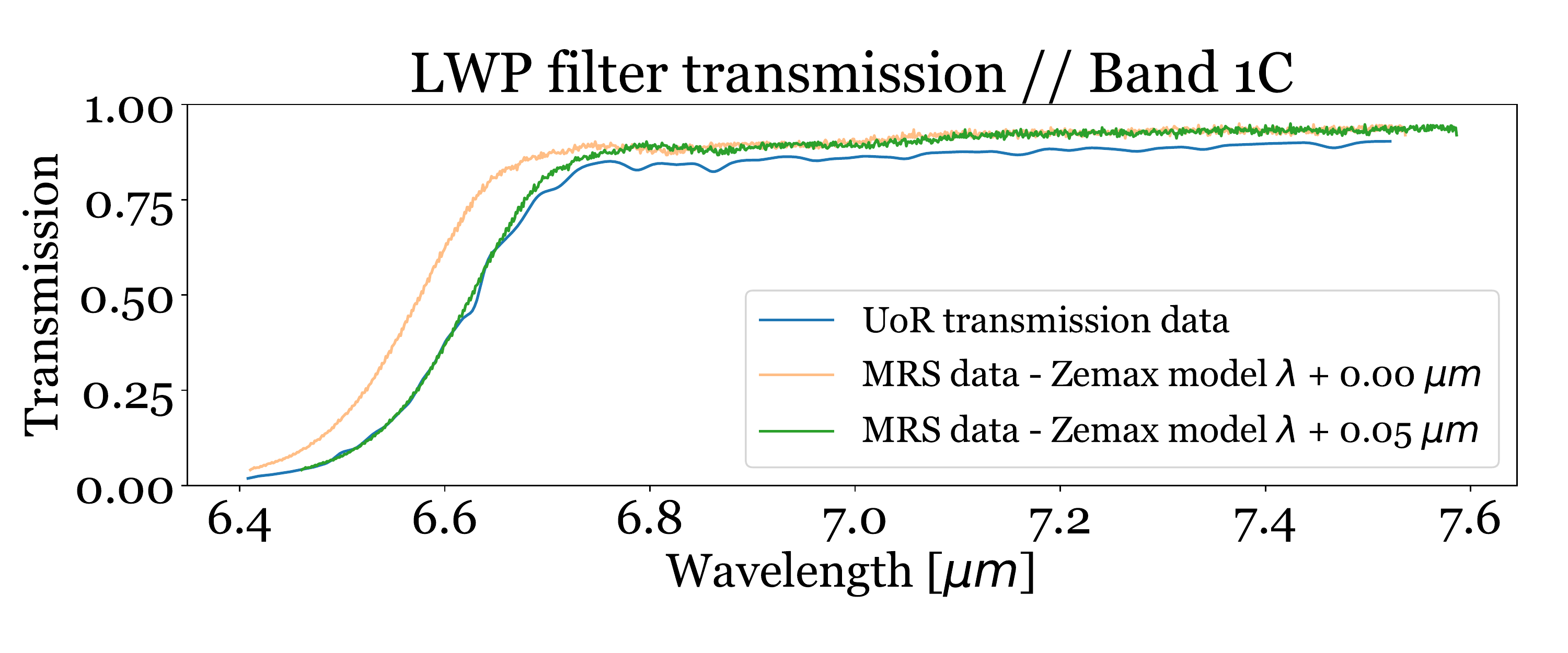}
\caption{Determination of global wavelength offset between UoR data and MRS data. The MRS data are wavelength calibrated using the Zemax optical model predictions.}
\label{fig:wavelength_offset_band1C}
\end{figure}

For band 1C the cut-off is not sharp enough to accurately determine the reference point wavelength-pixel pair. To address this we assumed that the wavelength scale of the Zemax model is correct and attempt to determine a global offset between the UoR data and the MRS data. Initially the MRS data are wavelength-calibrated based on the Zemax model. The global offset is computed by minimizing the transmission difference between the lab and MRS data below 6.6 $\mu m$, as shown in Fig.~\ref{fig:wavelength_offset_band1C}. An offset of +0.05 $\mu m$ is found. Translating this back to pixels, the cut-off of 6.6$\mu m$ is located at (x,y,$\lambda$) = (76pix,117.11pix,6.6$\mu m$). Sub-bands 2C and 4C were wavelength calibrated using the minima and maxima of the spectral features in the LWS and SWP. Sub-bands 2A, 2B, 4A, and 4B are wavelength calibrated similar to band 1C, namely by minimizing the difference in transmission between the UoR data and the MRS data.



\subsection{Overlapping spectral regions}
\label{overlap.sec}

Adjacent bands in the MRS show an overlap in wavelength coverage. This allowed us to 1) calibrate bands with no RPT defined, and 2) check consistency in the wavelength calibration of adjacent bands.

Band 1A, for example, has no spectral features to define a RPT (Table \ref{RPs}), but we had a RPT (and therefore a wavelength solution) for band 1B. In that case, we used the beating pattern of ET1 and ET2 to find the overlap region with band 1B. Once defined, we assigned the (x,y,$\lambda$) triplets to the etalon lines in the overlap region using the wavelength solution from band 1B. Then, we cross correlated the fitted peaks of the lines in wavelength (FTS) and pixels (FM data) for the rest of 1A, as we did with any other band, and continued the process normally to obtain a wavelength calibration for all slices in band 1A.

To check for the calibration consistency in different bands, we used the fact that, as shown in Table \ref{RPs}, all sub-bands except 4B and 4C include lines from at least two etalons. The order in which the lines of the different etalons appear, combine, and overlap ({\it beating pattern}), and the wavelengths of their centroids, must be the same in the overlaping region of two adjacent bands. Thus, by comparing the wavelength solution of two adjacent bands and the beating pattern of the etalons in the overlap regions, we confirmed that the wavelength solutions of all sub-bands, up to 4A were consistent with each other.

Channel 3, where there are no RPTs present, was calibrated using this method. In this case we had the advantage of using the overlaps with band 2C and 3A, and band 4A and 3C. In a similar manner, we calibrated 3B using the overlapping regions with 3A and 3C. Thus, we obtained a wavelength solution for all bands in channel 3, and also checked the consistency of the calibration of channels 2 and 4, as the solutions derived for channel 3 using wither of them had to be consistent.




\section{Relative Calibration}
\label{relative.sec}

\subsection{Expanding to a full solution}

\begin{figure}
\centering
\includegraphics[width=90mm]{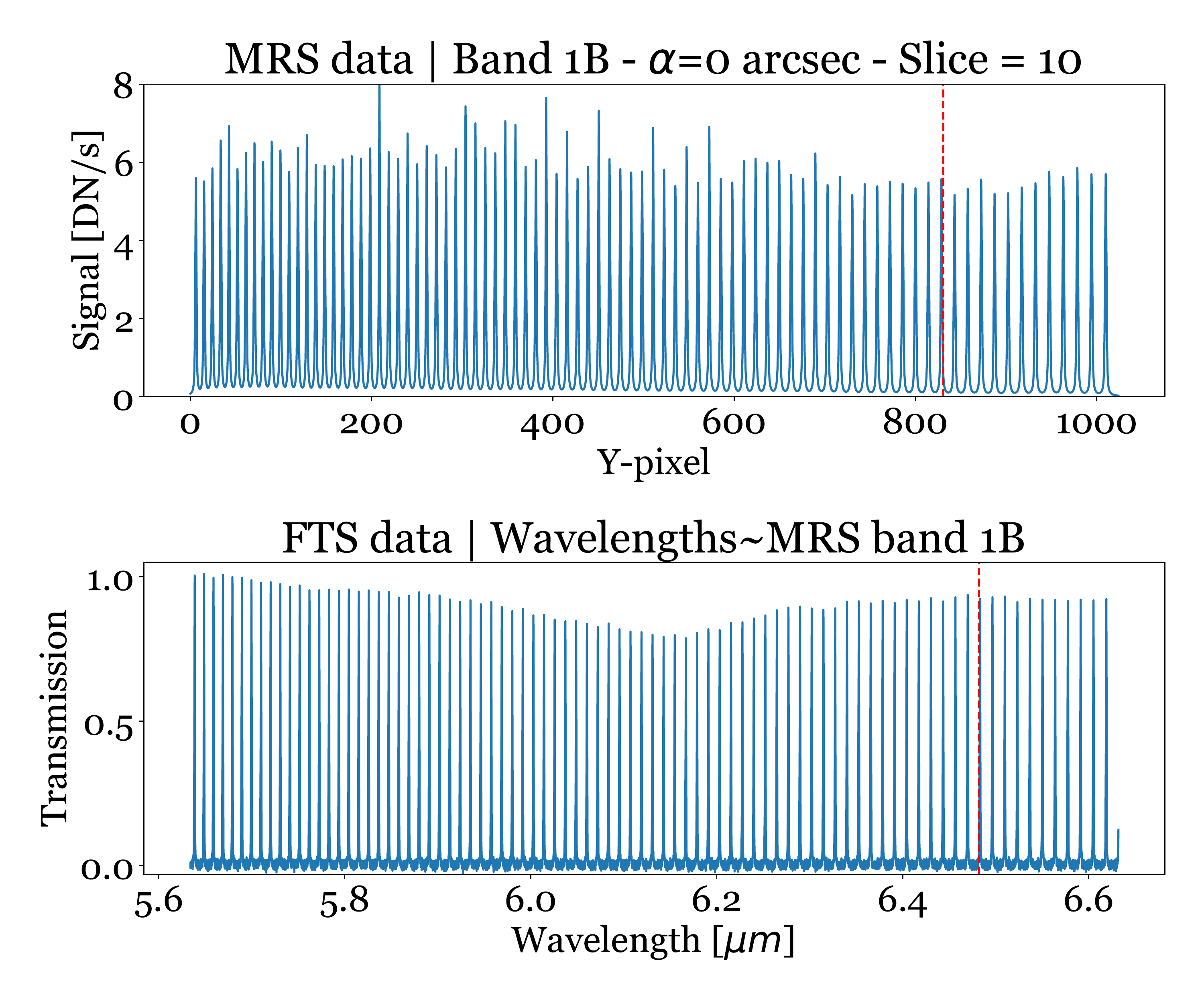}
\caption{A reference wavelength/pixel pair is used to match the MRS and FTS etalon spectra.}
\label{fig:fts_fm_matching}
\end{figure}

The identified (x,y,$\lambda$) triplets are used to identify the etalon lines around the RPT. By extrapolation, all the other etalon lines in the \isoa line are spectrally identified. Figure~\ref{fig:fts_fm_matching} shows the principle of matching the etalon lines in wavelength and in pixel space using the previously-determined reference point. The MRS line closest to the RPT is the same as the FTS line closest to the reference wavelength. As such, all the lines to the left and to the right of that line are also matched, and the pixel-to-wavelength mapping is complete for that \isoa line. 

Since the etalon is flood-illuminated, the etalon lines span the entire width of a slice. Once we had the central \isoa calibrated for a slice, we were then able to calibrate the entire slice following the \isol lines on the detector for that slice. The analysis of determining the reference point position, however, had to be repeated for the other slices.



\begin{figure}
\centering
\includegraphics[width=\columnwidth]{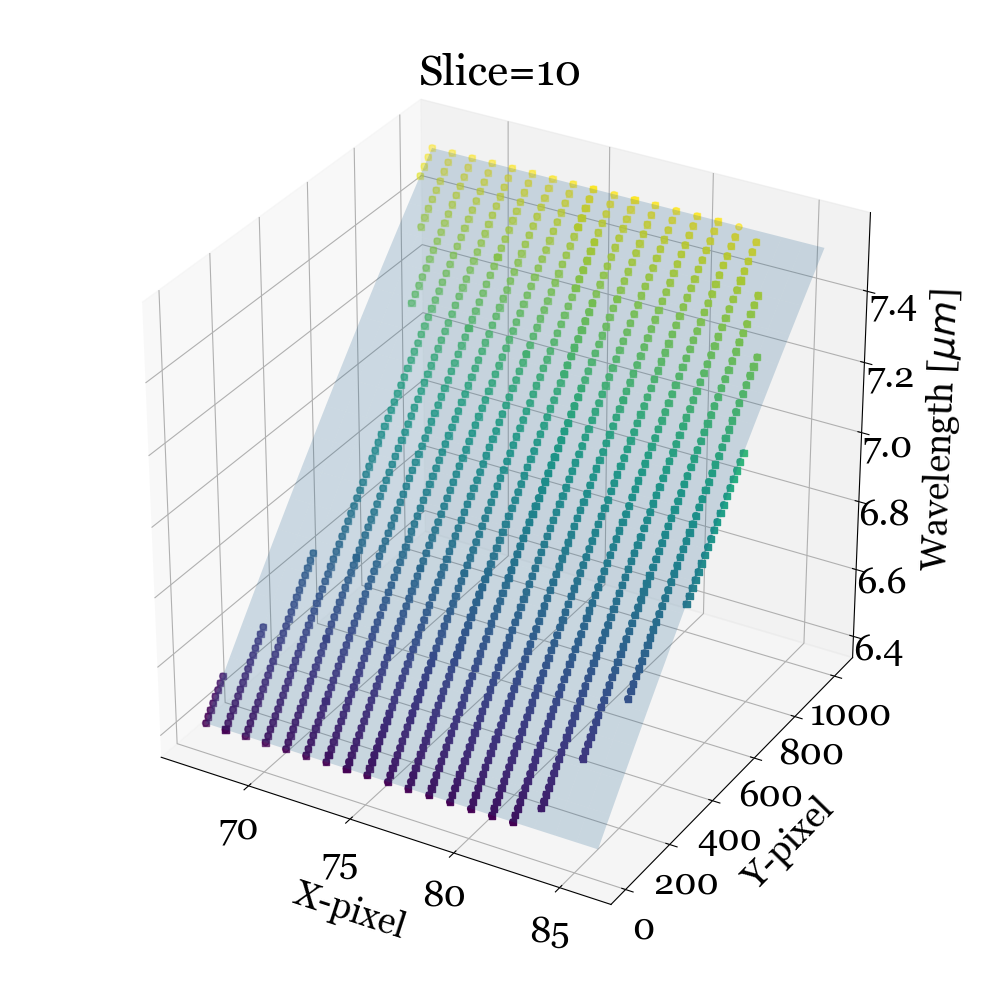}
\caption{
2D polynomial fit of the ET1 line centroids in slice 10 of MRS sub-band 1C. The scatter points show the ET1 centroids fitted on the detector (x,y) pixel space. The respective wavelength values matched using the FTS data are shown on the z-axis and in the color scale of the dots. X and y axes show the respective pixel coordinates on the detector plane. The light blue plane shows the fitted 2D polynomial.
}
\label{2d_fit}
\end{figure}

The wavelength assignation to etalon line peaks created a map of (x,y,$\lambda$) coordinate triplets for $\sim$10$\%$ of the pixels in the detector. 
The last step of the wavelength calibration was to fit a 2D polynomial function of { second order} to these triplets, yielding a parametrization of wavelength as a function of detector coordinates, $\lambda(x,y)$, for every point on the detector: 



\begin{equation}\label{{eq:polyval2d}}
\lambda_s(x,y)  = \sum_{i,j=0}^{N,N} K_{\lambda s} (i,j) \cdot (x - x_s)^j  \cdot y^i
\end{equation}

where s is the slice number, $N$ is the order of the polynomial, $K_{\lambda s}$ are the polynomial coefficients, and $x_s$ is the x-coordinate on the detector for the center of the slice in the middle of the detector (y=512). We produced one set of polynomials for each slice. 
Figure \ref{2d_fit} shows the resulting 2D fit of the (x,y,$\lambda$) triplets (i.e. the final, calibrated wavelength solution) of slice 10 of sub-band 1C.

\subsection{Refining the Relative Solution}


If all \isoa lines and slices of a sub-band were perfectly calibrated, a single etalon line should have the same transmission profile over wavelength anywhere on the detector. Thus, representing on one single plot the line profile of a line from all the spatial locations where it is present should produce the same transmission versus wavelength curve. Basically, the method is similar to "collapsing" all spatial direction on a single sub-band to produce one spectrum of the etalon, to produce a "super-sampled" spectrum with the different pixel phases along a given slice. 
With this method we are able to mitigate the uncertainties in defining the RPT in the different slices, as well as the uncertainties in the MRS etalon line centroid.

\begin{figure}
\centering
\includegraphics[width=90mm]{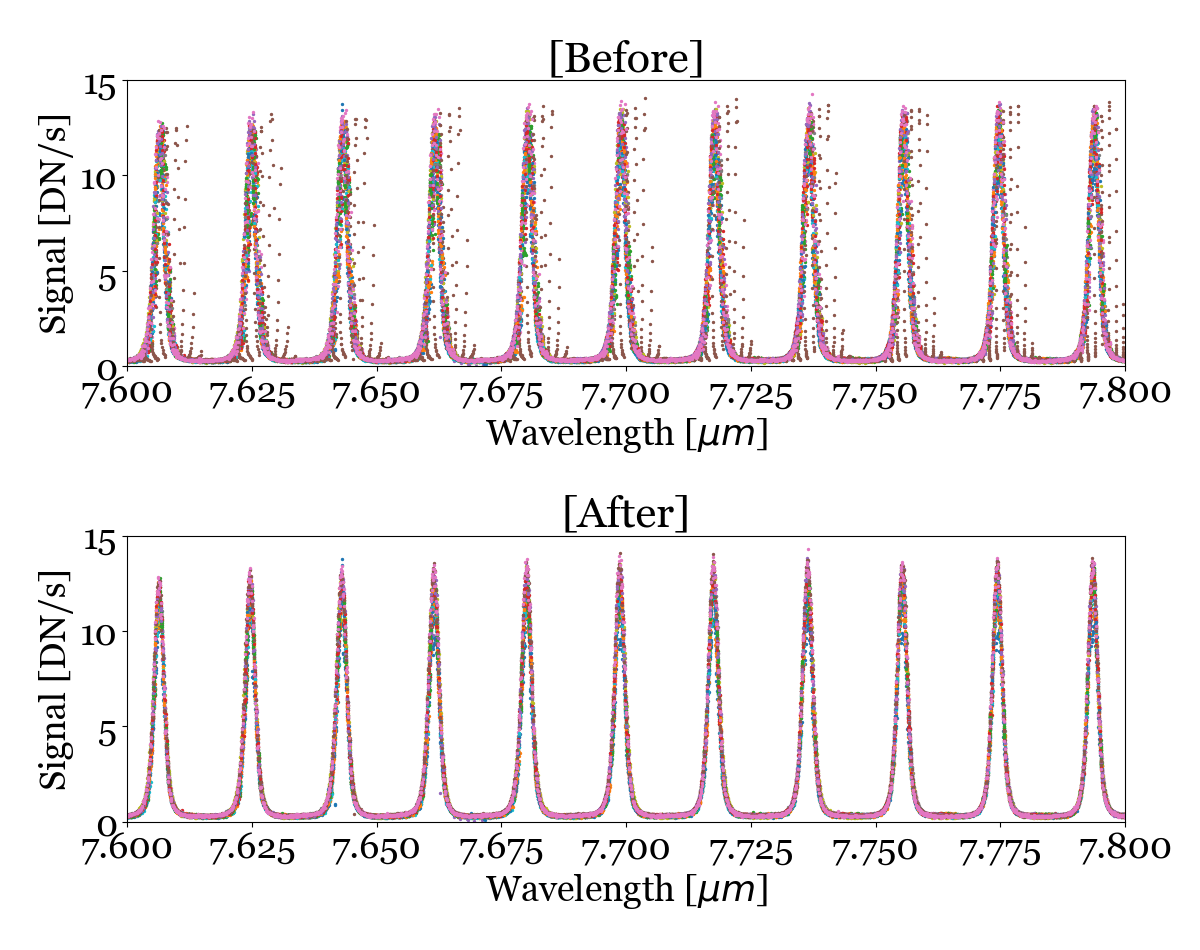}
\caption{Internal consistency check of the wavelength solution within a sub-band. The top panel shows a sub-optimal wavelength calibration of sub-band 2A. The misalignment of lines introduced by the uncertainties in the cross calibration of a few \isoa lines and slices is clearly seen as a shift to the right from the main "oversampled" lines. The bottom panel shows the resulting etalon spectrum once the uncertainties in the calibration have been addressed.}
\label{fig:internal_consistency_band2A_ET1A}
\end{figure}

The collapsing of the two spatial dimensions onto the spectral axis (i.e. wavelength space) is done by omitting the spatial coordinate of a given pixel in ($\alpha,\beta$) and plotting it based solely on its spectral coordinate ($\lambda$, in \mum). This effectively creates a point cloud in two dimensions, or equivalently, an oversampled 1d spectrum, as shown in Fig.~\ref{fig:internal_consistency_band2A_ET1A}. The top plot showcases a sub-optimal initial wavelength calibration of MRS band 2A. The calibration is impacted by the inaccurate determination of the reference point in some slices as well as the noise in the data. In the bottom plot we use the fact that all etalon lines should fall at the same wavelengths as an internal consistency check, correcting previous inaccuracies. Producing a similar plot as Fig.~\ref{fig:internal_consistency_band2A_ET1A} for all spectral sub-bands immediately shows if there are inconsistencies in the wavelength calibration within any sub-band. The method works as follows:
\begin{enumerate}
    \item We define a reference slice on the detector. The reference slice is chosen on the basis of a plot similar to that shown in the top plot of Fig.~\ref{fig:internal_consistency_band2A_ET1A}. The criterion is that the oversampled spectrum resulting from the reference slice should not deviate from the majority of the oversampled spectra in the other slices. The discrepancy is very clear as only a couple of slices show large offsets.
    \item We use the new wavelength solution to define a reference etalon spectrum in wavelength space. The reference spectrum is an oversampled spectrum made by compiling all the pixels in the reference slice. This oversampled spectrum is used as the absolute wavelength calibration reference. 
    \item We fit the etalon lines of the oversampled MRS spectrum in the reference slice using pseudo-Voigt profiles.
    \item We fit the spectra extracted from all the other slices.
    \item We substitute the wavelength centroids of the etalon lines in the other slices (same for all \isoa lines in the slice) by the centroid of the etalon line that is found closest in the reference spectrum.
    \item We re-fit the 2D polynomial to the boot-strapped etalon line centroids.
\end{enumerate}

By recognizing that the same feature in different \isoa lines and slices should have the same wavelength centers, the spectral calibration is improved drastically and allowed us to achieve the MIRI-required spectral precision down to 1/10th of a resolution element. Table \ref{wavtab} shows the resulting MRS wavelength coverage based on our ground calibration.


\section{Spectral Resolving Power}
\label{resolution.sec}

\begin{figure*}
\centering
\includegraphics[width=\hsize]{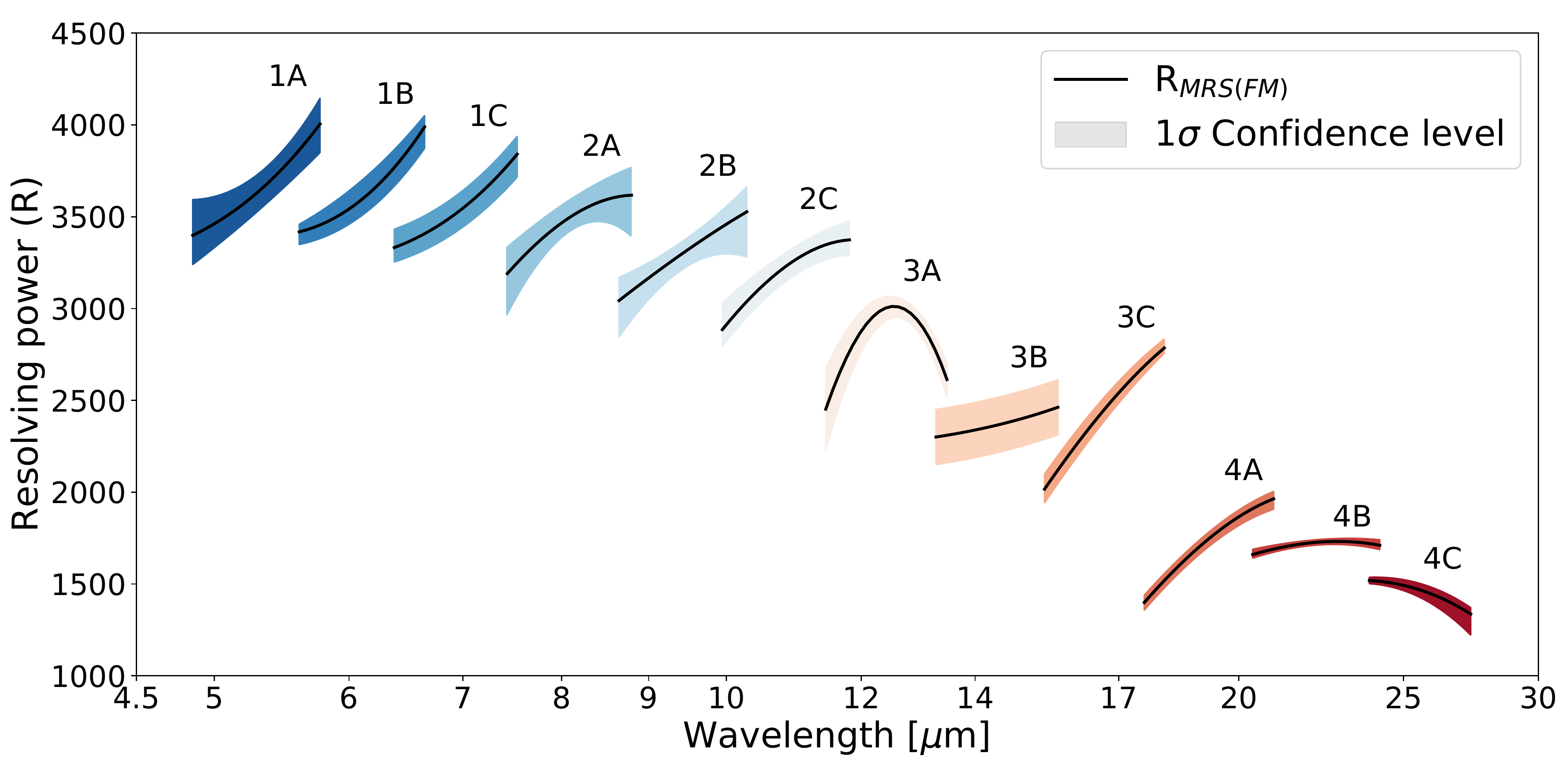}
\caption{{Best estimation of the of the MRS resolving power, based on ground data. Black line: median MRS resolving power. Colored area: 1$\sigma$ confidence level. R$_{MRS(FM)}$ is really a lower limit of the expected MRS resolving power, which could be probably up to a 10\% higher (see text for details).}}
\label{fig1_resol}
\end{figure*}

\begin{table*} 
\caption{\label{wavtab} Calibrated spectral characteristics of the MRS derived from ground data.}
\centering
\begin{tabular}{ccccccc}
\hline\hline
MRS    &MRS &Wavelength coverage & Pixel scale$^*$ & R best estimate & R$_{(FM+FTS)}^{**}$ & Spectral sampling\\
Channel&Band& (\mum) & \mum/pixel & R$_{MRS(FM)}$ & R$_{MRS(max)}$ & LSF FWHM (pixels) \\
\hline
   & SHORT (A)  &   4.885 -- 5.751  & 0.0009 & 3400 -- 4000 & 3630 & 1.70 -- 1.70 \\
1 & MEDIUM (B)  &   5.634 -- 6.632  & 0.0010 & 3420 -- 3990 & 3620 & 1.69 -- 1.71 \\
    & LONG (C)  &   6.408 -- 7.524  & 0.0011 & 3330 -- 3840 & 3590 & 1.77 -- 1.80 \\
\hline 
  & SHORT (A)   &   7.477 -- 8.765  & 0.0013 & 3190 -- 3620 & 3520 & 1.86 -- 1.92 \\
2 & MEDIUM (B)  &   8.711 -- 10.228 & 0.0015 & 3040 -- 3530 & 3290 & 1.86 -- 1.89 \\
    & LONG (C)  &  10.017 -- 11.753 & 0.0017 & 2890 -- 3374 & 3250 & 2.04 -- 2.05 \\
\hline
  & SHORT (A)   &  11.481 -- 13.441 & 0.0019 & 2450 -- 3010 & 3010 & 2.45 -- 2.33 \\
3 & MEDIUM (B)  &  13.319 -- 15.592 & 0.0022 & 2300 -- 2460 & 2370 & 2.61 -- 2.86 \\
    & LONG (C)  &  15.400 -- 18.072 & 0.0026 & 2020 -- 2790 & 2470 & 2.92 -- 2.48 \\
\hline
  & SHORT (A)   &  17.651 -- 20.938 & 0.0032 & 1400 -- 1960 & --  & 3.33 -- 3.93  \\ 
4 & MEDIUM (B)  &  20.417 -- 24.220 & 0.0037 & 1660 -- 1730 & --  & 3.31 -- 3.77  \\ 
    & LONG (C)  &  23.884 -- 28.329 & 0.0043 & 1340 -- 1520 & --  & 4.11 -- 4.29  \\ 
\hline
\end{tabular}
\tablefoot{\tablefoottext{*}{Calculated as wavelength coverage / 1024 pixels.}\tablefoottext{**}{ R$_{MRS(FM+FTS)}$ given for central wavelength of the sub-band. It could not be determined for channel 4 due to artificial widening of the etalon lines introduced by the FTS. See text for details.}}
\end{table*}

The resolving power (R) is defined as $\lambda/\Delta\lambda$, where $\Delta\lambda$ is the minimum distance to distinguish two features in a spectrum. The shape of the etalon emission lines seen by the MRS during the FM campaign is well defined by a Voigt profile (see Section \ref{obs.sec}). It is the result of the convolution of the intrinsic etalon line described by Lorentzian profile, and the MRS Line Spread Function (LSF), that has been approximated to a Gaussian. {To avoid dealing with deconvolution artifacts, we use an approximate relation linking the widths of a Voigt, a Lorentzian, and a Gaussian profiles \citep{empirical_voigt_deconvolution}:}

\begin{equation}\label{eq:deconvolved_fwhm}
\begin{aligned}
    FWHM_{FM} \approx  0.5346 \cdot FWHM_{ET} +  \\
                      + \sqrt{ 0.2166 \cdot FWHM_{ET}^2 + FWHM_{MRS}^2 }
\end{aligned}
\end{equation}

{where, FWHM$_{FM}$ is the FWHM of the etalon line in the FM data (Voigt profile), FWHM$_{ET}$ is the FWHM of the etalon line in the FTS results (Lorentzian profile), and FWHM$_{MRS}$ is the FWHM of the MRS LSF, which we assume can be described by a Gaussian profile. Eq. \ref{eq:deconvolved_fwhm} have an accuracy of 0.02\%.}

If we define the FWHM$_{MRS}$ as the minimum distance to distinguish two features in the spectrum ($\Delta\lambda$), the MRS resolving power can be defined as:  
\begin{equation}\label{eq:Rmrs}
    R_{MRS(FM+FTS)} = \frac{\lambda}{FWHM_{MRS}}
\end{equation}
where $\lambda$ is the central wavelength of the etalon line and FWHM$_{MRS}$ is obtained from Eq. \ref{eq:deconvolved_fwhm}.

{The determination of the intrinsic etalon line profile is significantly complex due to the uncertainties on the FTS measurements. The etalons were illuminated by a quasi-collimated beam generating an asymmetric line profile (Section \ref{obs.sec}). When the FTS etalon line is corrected by the cone angle effect, the FWHM is reduced up to 50$\%$ from the observed FTS values. Additionally, FTS FWHM values do not agree with the ones simulated by INTA, taking into account all the fabrications parameters achieved by the manufacturer. The simulated FTS FWHMs of ET1 are lower by a factor of 2-3, however etalons ET2, ET3, and ET4 present higher FTS FWHMs by a factor of 1.2-1.1, 1.8-1.1, and 5.6-2.2, respectively. If the intrinsic FWHM of the etalon emission lines are not consistent with the FTS determinations, the MRS resolving power could be overestimated. Then, assuming that the intrinsic width of the etalon lines is negligible (i.e. FWHM$_{ET}$=0), we calculated a lower limit to the MRS resolving power:}
\begin{equation}\label{eq:Rmrsmin}
    R_{MRS(FM)} = \frac{\lambda}{FWHM_{FM}}
\end{equation}    

We used the wavelength calibration results, together with the intrinsic width of the etalon lines obtained from the FTS data, to derive both estimations of the MRS resolving power, R$_{MRS(FM+FTS)}$ and R$_{MRS(FM)}$. The estimation was done for each etalon line present in each slice and sub-band. {To estimate the average resolving power of each sub-band as a function of wavelength, we divided the sub-band into seven wavelength bins to calculate the median, and the $\pm 1\sigma$ confident level. The median and confidence level of all bins have been fitted with a polynomial function of second order. The confidence level gives the variations as a function of location throughout the MRS field \citep{Wells15, Law20} for each sub-band. These variations will be measured in detail during the commissioning and in-flight calibration campaigns.}

{R$_{MRS(FM+FTS)}$ tends to overestimate the MRS resolving power at longer wavelengths due to the large uncertainty in the FTS data. We found that the intrinsic width of the etalon emission lines from the FTS data is  higher than the width in the simulations provided by INTA ( based on the fabrication parameters given by the manufacturer) in channels 2 and 3, and even higher than the widths in the FM data in channel 4. This suggests that the FTS is artificially widening the etalon emission lines on those channels,  overestimating therefore the MRS resolving power. However, that is not the case for channel 1, where the FTS data are consistent with the INTA simulations, and the R$_{MRS(FM+FTS)}$  represents the best estimate of the MRS resolving power. In this case, the R$_{MRS(FM+FTS)}$ is in average 10\% higher than the R$_{MRS(FM)}$.}

{Figure \ref{fig1_resol} shows the R$_{MRS(FM)}$ which, based on the discussion above, is the best  achievable estimation of the MRS resolving power with the current ground data. R$_{MRS(FM)}$ is really a lower limit, as the intrinsic width of the etalon emission line has been assumed negligible, but it is consistent with the estimations based on Zemax optical models \citep{Wells15}, suggesting that the real resolving power of the MRS is close to R$_{MRS(FM)}$. Extrapolating the R$_{MRS(FM+FTS)}$ results for channel 1 to longer channels, the R$_{MRS(FM)}$ estimates should represent the MRS resolving power within a 10\% shift to higher values. The final resolving power of the MRS will be further investigated and updated during commissioning and in-flight calibration campaigns.}


\section{Caveats}
\label{caveats.sec}

\subsection{Etalon collimation}

The main source of uncertainties in our the wavelength calibration analysis are the etalons. The distance between the peaks of two adjacent lines is given by the free spectral range (FSR):

$$
FSR = \frac{\lambda^2_0}{2nd~cos(\theta)}
$$

where $\lambda_0$ is the wavelength of the peak of the line, n is the refraction index of the medium in the etalon cavity, d is the width of the etalon spacing, and $\theta$ is the angle of incidence of the illumination beam. The FTS measurements described above were done at 80K while the FM data was obtained at operating temperature of 40K. That difference in temperature affects the size of the spacing ($d$) and thus the distance between the peaks of the etalons. That means the distance between etalon lines (in micron) in the FTS data is different to the distance between etalon lines in FM data (in micron). To estimate the error introduced by the temperature difference in the etalon spacing, we used literature information \citep{Browder69, Browder72, Smith75} on the thermal expansion coefficient for the etalon materials (ZnSe for ET1,2, and 3, and CdTe for ET4). Our estimations yield a negligible variation of 4--8\% of a pixel for all bands \citep{wavecal_spie2010, TesisRafael}.

The transmission of the 4 etalons were measured with a quasi-collimated beam at normal incidence. Even though the uncertainties of these measurements are not available, we dealt with them with the pseudo Voigt profile and a fit of the incidence angle in the fit of the etalon line shape and position. Our fits show that the divergence angle of the beam is $\psi<1\deg$. This angle created a small shift ({$\lesssim0.001$\mum}) on the FTS data, as $\theta$ in the FSR equation is in this case all angles from $-\psi/2$ to $+\psi/2$ for each etalon line. By fitting this divergence angle and how it affects the beam, we corrected the positions of the FTS etalon lines in wavelength before using them to obtain for the MRS wavelength dispersion solution. 



Another source of uncertainty is the definition of the RPTs for the cut-off filters and dichroics. The fitting methods used to define those RPTs have uncertainty $\lesssim$8 pixels, which produces offsets in wavelength for the whole band where the RPT is used. The beating pattern used in the overlap regions should correct for small variations between bands. However, there is room for a total offset in wavelength from the ground calibration to the real one.

Both the total offset in wavelength (if present) and the variations in pixel scale (\mum/pixel) described above will be accounted for and corrected during commissioning, as we will use as RPTs known emission lines of the targets.


The main caveat on the resolving power determination with the ground data is the unknown intrinsic profile of the etalon emission lines. The lack of uncertainties of FTS etalon measurements and the effect of the non-collimated beam, together with the possible effect on the different temperature used in the FTS and FM measurements do not allow us to well characterize the intrinsic etalon lines profile. The uncertainties presented in Fig. \ref{fig1_resol} are directly related with the uncertainty on the determination of the intrinsic etalon line profile. It could be up to 40$\%$ in channels 2 and 3, and in channel 4 only a lower limit could be given. The commissioning activities will allow to reduce the current uncertainty with the observations of the emission lines of sources well characterized in the literature.

\subsection{Across-Slice Wavelength Correction}




The MRS spectra are expected to show a small shift in the wavelength direction of the spectral image formed on the detector, which arises due to (sub slice width) spatial offsets of a compact target in the across-slice direction. This is a known effect of slicer-based IFUs (as opposed to fibers or lenslets which scramble the light and avoid degeneracy between the cross-slice and dispersion directions). The spatial offset gives rise to a shift in the centroid of the illumination pattern across the slice; this then translates into a shift in the dispersion direction of the output spectrum. 
The shift, which has a {typical value} of up to 20 \kms, is only seen from compact targets and has an amplitude which varies with the source’s across-slice location. For extended sources, the effect from each \isoa averages out over the slice. Thus, this effect is only present when an individual source is extracted (the generation of a single spectrum from the multiple spectral elements measured using the MRS’ multiple spatial samples).

The optimal spectral extraction routine for point sources in the MRS pipeline \citep{MRSpipeline} has a step dedicated to correct the wavelength scale for multiple adjacent slices. 
With the pipeline step correction we produce a spectral correction  accurate to better than $\pm$0.02 spectral resolution elements.

\section{In-Flight Calibration Strategy}
\label{flight.sec}

\subsection{Overview}

To optimize our use of time during commissioning, we will just confirm and adjust the wavelength calibration rather than re-deriving it from scratch. The main goal of the commissioning is then to check the ground calibration and update or improve it where needed. For the MRS spectral characterization, the strategy will be to observe several targets with multiple, bright, unresolved emission lines in the 12 sub-bands. These targets are chosen to be visible during the expected observing time for this commissioning activity (towards the end of the JWST commissioning campaign). 

The target list includes unresolved ("point") sources, as well as sources that fill the MRS FOV ("extended"). We need to observe point sources to obtain an accurate estimation of the resolving power of the MRS. However, the point source data will be taken for very few slices (and positions in those slices) in the MRS detectors. We will therefore use the extended source data to extrapolate the wavelength calibration given by the point source 
to the whole detector. 


The point source data will also be used to characterize the across slice wavelength variations, taking advantage of the different slice locations of the target in each of the dither pointings. Discrepancies between the ground calibration and the known wavelength values of the emission lines (from atomic measurements) will be used to update the ground calibration where needed. This update will be done in a similar way as the ground calibration, where the emission lines of the commissioning targets will be used as RPTs.  
We will also use the overlap regions of the spectral bands to cross check the consistency of the in-flight calibration

The density of spectral lines in the target sources should be sufficient to establish a good dispersion relation for all spectroscopic modes. Since many of the lines will be unresolved (the best velocity resolution $\Delta v$, equivalent to $FWHM_{MRS}$ in wavelength space, is about 80 km/sec, see Section \ref{resolution.sec}), the observed line widths can also be used to update the spectral resolution. Some lines are expected to be closely spaced and can also provide an estimates of the spectral resolution. 


\subsection{Target selection}

\begin{figure}
\centering
\includegraphics[width=90mm]{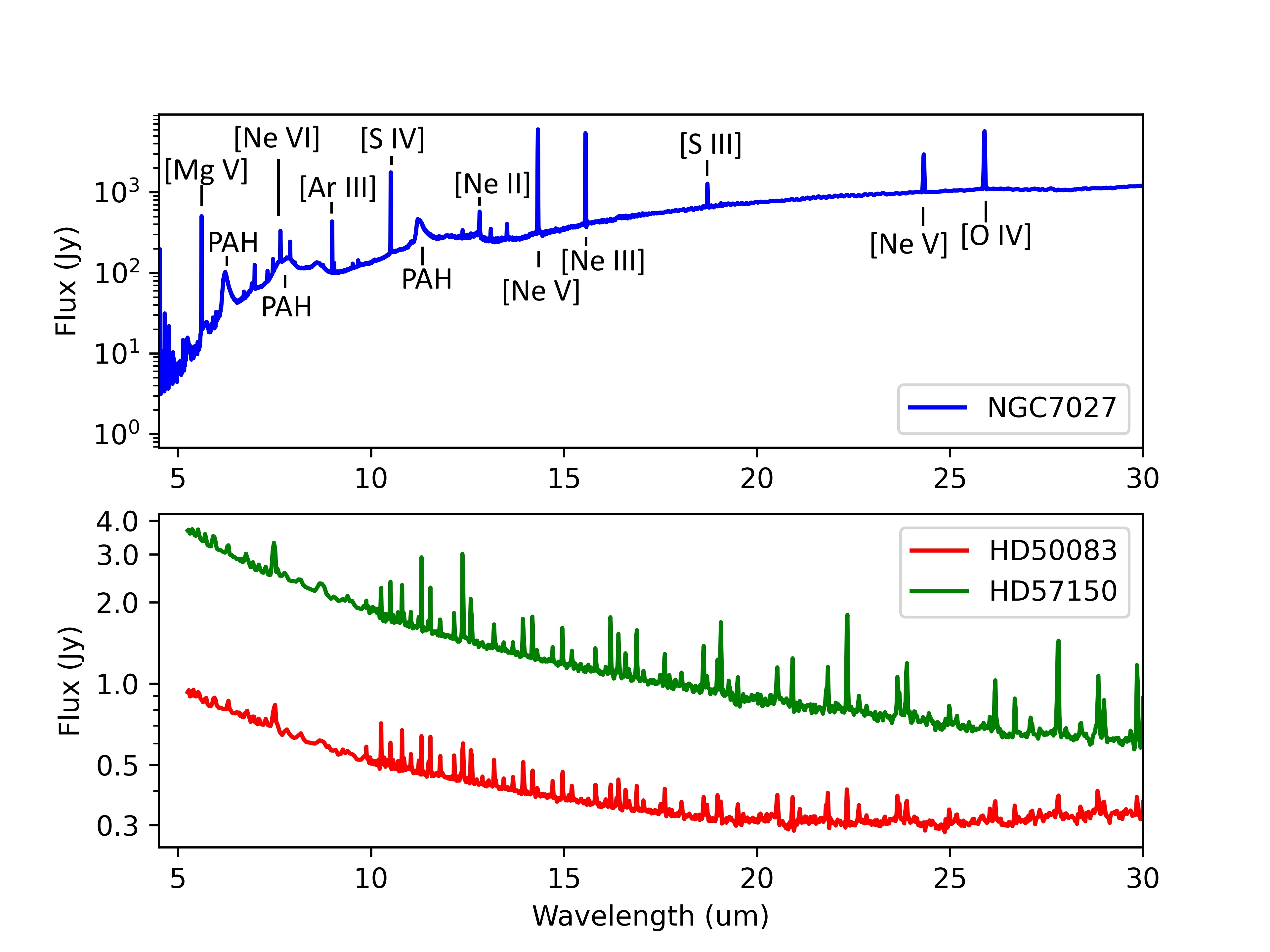}
\caption{Top: ISO-SWS spectrum of NGC7027. This spectrum covers a much larger field of view than the MRS. Thus, the fluxes expected in the  MRS data should be much lower. Bottom: IRS spectra of HD57150 and HD50083.} 
\label{comm_pn}
\end{figure}

Figure \ref{comm_pn} show the Spitzer InfraRed Spectrograph (IRS) and ISO Short Wave Spectrometer (SWS) spectra of the targets for this commissioning activity. Below, we list a few of the properties of the targets selected. 

{\it Point sources}: HD57150 (K=4.516 mag, 2MASS) and HD50083 (K=6.292 mag, 2MASS) are bright Be stars with plenty of bright, unresolved emission lines in the MRS wavelength coverage. HD57150 is the brightest target in the list, and very close to the saturation limit given by the JWST Exposure Time Calculator (ETC).

We recommend to observe both stars. Indeed, the available MIR spectra for those sources have a much lower spectral resolution than MIRI-MRS’. Consequently, the MRS spectra could unveil unexpected line structures, asymmetries and velocities, resolved emission from discs, higher fluxes than expected (risking saturation), etc., invalidating the use of one of the sources for the purpose of wavelength calibration.


{\it Extended source}: NGC7027: a close-by planetary nebula with many bright, unresolved emission lines. For the ETC estimations, we used the ISO-SWS spectrum  to create a 20"x20" source (the ISO-SWS average field). For this commissioning activity, we will use a pointing in the ring of the nebula, where the emission lines are generated. We are also considering observations of NGC6302, which shows similar emission line fluxes, but a fainter continuum. 



{ The IRS spectra of the stars shows that all sub-bands will include several emission lines to update the wavelength calibration. There is no high-resolution spectra, at MIRI wavelength ranges for this objects, so there is a slight uncertainty on the sharpness of the lines. Some of these could be resolved by the MRS. That is not a problem for the wavelength calibration, but could affect the resolving power estimations. We do have, however, enough lines all over the MIRI range in the stars and the planetary nebula to guarantee unresolved emission features. Some regions of the nebula are expected to show intrinsic velocities, which will yield shifted line centroids. We plan to address those shifts by comparing the spectra, and the wavelength calibration produced by the three targets independently. The cross calibration among adjacent bands should also correct for isolated shifts in the lines. Also, we are currently working on MIRISim \citep{mirisim} simulations of the targets with the flight pipeline to improve the analysis strategy if needed \citep[e.g.,][]{Alvarez-Marquez+19}. Taking all these into account, and based in the instrument models and extensive analysis of the data, we expect variations between the ground and on-orbit wavelength calibration below 1-5 pixels at most. Our results over the years suggest that this variation will most likely happen as a wavelength offset, without varying significantly the dispersion (arsec/pixel) on the spectral direction. If we needed to update the dispersion, we can do it with the emission lines from the celestial targets (three per sub-band should be enough), and also reuse the etalon transfer curves to fine tune it, based on the RPs obtained on orbit.}

The observed data will be corrected for { radial velocity (v = 9 \kms\ for NGC 7027)} and heliocentric velocity before assigning wavelengths. Also, the FM data shows no skewness dependence on position on the FOV. Even though we don't expect a skewness dependency on FOV position, we will need to confirm its absence using the extended source commissioning data.

{ The MIRI optics were designed with dithering in mind. On non-dithered observations of point sources, we observe clearly the effect of the pixel/slice phase. The convolution of the input PSF with the top-hat pixel/slice response function results in an effective combined PSF profile (FWHM and minor/major axis ratio) that vary strongly within the field of view depending on whether a source is located in the middle of a sampling element or between two elements \citep[see, e.g., Fig 6 of][for a similar effect on fiber-type IFUs]{Law15}. The dithering strategy is designed to mitigate this sampling issue and provide as uniform a PSF profile as possible. While this is an interesting discussion in its own right, it is well beyond the scope of this paper.}

{ We also expect some variations in the resolution with FOV position, as is the case in most IFUs \citep[e.g.,][]{Law20}. These variations will be characterised during commissioning. They are not expected to be significant for most science cases, as most of the deviations in R we see in the ground data are mainly caused by the uncertainties in the etalon calibration and FTS measurements.}


\section{Summary}
\label{summary.sec}

The ground test campaigns, carried under flight conditions, produced a large amount of the data to calibrate the MRS in detail. Concerning the spectral characterization, the MTS provided all tools necessary to measure the spectral dispersion and resolution of the MRS. The wave-pass filters on the MTS, and the cross-dichroic configurations of the MRS showed clear spectral features at known wavelengths, providing RPTs to establish an absolute wavelength scale. Four MTS etalons (each optimized for each MRS channel) were designed to produce enough unresolved transmission lines to measure the wavelength values for enough pixels all across the detector. A 2D polynomial fit per slice completed the wavelength calibration for the whole detector. Four MRS bands had no RPTs defined. 

By construction, all adjacent MRS bands have an overlap in wavelength. Combining the etalon transmission curves, and the absolute wavelength calibration of bands with RPTs, we were able to extrapolate the absolute calibration of the RPT calibrated bands to those with no RPTs. The overlap regions were also used to cross-check the calibration of adjacent bands. An oversampling method was used to refine the wavelength calibration down to the required precision of 1/10 of a resolution element.

The etalon transmission curves were also used to estimate the resolving power of the 12 bands. Some problems on the ground measurements produced an uncertainty of ~10$\%$ in the final values. Taking into account this uncertainty, our results suggest a better resolving power than the initial MIRI/MRS requirements. Table \ref{wavtab} shows the wavelength coverage and resolving power based on the ground calibration. 

We also presented the commissioning strategies that will be followed to update the spectral characterisation of the MRS, and confirm the resolving power measured. The target selection was for the commissioning campaign was also discussed the different targets proposed. These targets can further be used for the calibration campaigns during the routine operations of the JWST


\begin{acknowledgements}
       We thank the anonymous referee for her/his very useful recommendations. 
       
       A.L. acknowledges the support from Comunidad de Madrid through the Atracci\'on de Talento grant 2017-T1/TIC-5213. 
       
       A.L and J.A.M. acknowledge support from the Spanish State Research Agency (AEI) under grant numbers ESP2017-83197, and PID2019-106280GB-I00, and the support from CSIC through ILINK grant LINKB20003. This research has been partially funded by the Spanish State Research Agency (AEI) Project MDM-2017-0737 Unidad de Excelencia Mar\'ia de Maeztu - Centro de Astrobiolog\'ia (INTA-CSIC).
       
       Ioannis Argyriou thanks the European Space Agency (ESA) and the Belgian Federal Science Policy Office (BELSPO) for their support in the framework of the PRODEX Programme. \\

        The work presented is the effort of the entire MIRI team and the enthusiasm within the MIRI partnership is a significant factor in its success. MIRI draws on the scientific and technical expertise of the following organisations: Ames Research Center, USA; Airbus Defence and Space, UK; CEA-Irfu, Saclay, France; Centre Spatial de Liége, Belgium; Consejo Superior de Investigaciones Científicas, Spain; Carl Zeiss Optronics, Germany; Chalmers University of Technology, Sweden; Danish Space Research Institute, Denmark; Dublin Institute for Advanced Studies, Ireland; European Space Agency, Netherlands; ETCA, Belgium; ETH Zurich, Switzerland; Goddard Space Flight Center, USA; Institute d'Astrophysique Spatiale, France; Instituto Nacional de Técnica Aeroespacial, Spain; Institute for Astronomy, Edinburgh, UK; Jet Propulsion Laboratory, USA; Laboratoire d'Astrophysique de Marseille (LAM), France; Leiden University, Netherlands; Lockheed Advanced Technology Center (USA); NOVA Opt-IR group at Dwingeloo, Netherlands; Northrop Grumman, USA; Max-Planck Institut für Astronomie (MPIA), Heidelberg, Germany; Laboratoire d’Etudes Spatiales et d'Instrumentation en Astrophysique (LESIA), France; Paul Scherrer Institut, Switzerland; Raytheon Vision Systems, USA; RUAG Aerospace, Switzerland; Rutherford Appleton Laboratory (RAL Space), UK; Space Telescope Science Institute, USA; Toegepast- Natuurwetenschappelijk Onderzoek (TNO-TPD), Netherlands; UK Astronomy Technology Centre, UK; University College London, UK; University of Amsterdam, Netherlands; University of Arizona, USA; University of Bern, Switzerland; University of Cardiff, UK; University of Cologne, Germany; University of Ghent; University of Groningen, Netherlands; University of Leicester, UK; University of Leuven, Belgium; University of Stockholm, Sweden; Utah State University, USA. A portion of this work was carried out at the Jet Propulsion Laboratory, California Institute of Technology, under a contract with the National Aeronautics and Space Administration.
        
        We would like to thank the following National and International Funding Agencies for their support of the MIRI development: NASA; ESA; Belgian Science Policy Office; Centre Nationale D'Etudes Spatiales (CNES); Danish National Space Centre; Deutsches Zentrum fur Luft-und Raumfahrt (DLR); Enterprise Ireland; Ministerio De Economiá y Competividad; Netherlands Research School for Astronomy (NOVA); Netherlands Organisation for Scientific Research (NWO); Science and Technology Facilities Council; Swiss Space Office; Swedish National Space Board; UK Space Agency.
        
        This research has made use of ESASky, developed by the ESAC Science Data Centre (ESDC) team and maintained alongside other ESA science mission's archives at ESA's European Space Astronomy Centre (ESAC, Madrid, Spain).
        
        We take this opportunity to thank the ESA JWST Project team and the NASA Goddard ISIM team for their capable technical support in the development of MIRI, its delivery and successful integration.
\end{acknowledgements}

\bibliographystyle{aa} 
\bibliography{Bibliography.bib} 

\end{document}